\documentclass[a4paper,fleqn,usenatbib]{mnras}

\usepackage[T1]{fontenc}
\usepackage{ae,aecompl}

\usepackage{natbib}
\usepackage{graphicx}
\usepackage{enumitem}
\usepackage{calc}
\usepackage{soul}
\usepackage{lipsum,adjustbox}
\usepackage{subcaption}
\usepackage{tikz}
\usepackage{amsmath}	
\usepackage{amssymb}
\usepackage{wasysym}
\usepackage{txfonts}
\usepackage{array}
\usepackage{threeparttable}
\usepackage{float}

\title[Finding solar-like oscillators in \textit{Kepler} LC data]{A Search for Red Giant Solar-like Oscillations in All \textit{Kepler} Data}

\author[Hon et al.]{Marc Hon$^{1}$\thanks{E-mail: mtyh555@uowmail.edu.au},
Dennis Stello$^{1,2,3}$,
Rafael A. Garc\'ia$^{4,5}$,
Savita Mathur$^{6,7,8}$,
\newauthor
Sanjib Sharma$^{2}$,
Isabel L. Colman$^{2,3}$,
and Lisa Bugnet$^{4,5}$
\newauthor
\\
$^{1}$School of Physics, The University of New South Wales, Sydney NSW 2052, Australia\\
$^{2}$Sydney Institute for Astronomy (SIfA), School of Physics, University of Sydney, NSW 2006, Australia\\
$^{3}$Stellar Astrophysics Centre, Department of Physics and Astronomy, Aarhus University, Ny Munkegade 120, DK-8000 Aarhus C, Denmark\\
$^{4}$IRFU, CEA, Universit\'e Paris-Saclay, F-91191 Gif-sur-Yvette, France\\
$^{5}$Universit\'e Paris Diderot, AIM, Sorbonne Paris Cit\'e, CEA, CNRS, F-91191 Gif-sur-Yvette, France\\
$^{6}$Instituto de Astrof\'isica de Canarias, E-38200, La Laguna, Tenerife, Spain\\
$^{7}$Universidad de La Laguna, Departamento de Astrof\'isica, E-38205, La Laguna, Tenerife, Spain\\
$^{8}$Space Science Institute, 4750 Walnut Street Suite 205, Boulder, CO 80301, USA
}

\date{Accepted XXX. Received YYY; in original form ZZZ}

\pubyear{2018}

\begin{document}
\label{firstpage}
\pagerange{\pageref{firstpage}--\pageref{lastpage}}
\maketitle

\begin{abstract} 
The recently published \textit{Kepler} mission Data Release 25 (DR25) reported on $\sim$197,000 targets observed during the mission. Despite this, no wide search for red giants showing solar-like oscillations have been made across all stars observed in \textit{Kepler}'s long-cadence mode. In this work, we perform this task using custom apertures on the \textit{Kepler} pixel files and detect oscillations in 21,914 stars, representing the largest sample of solar-like oscillating stars to date. We measure their frequency at maximum power, $\nu_{\mathrm{max}}$, down to $\nu_{\mathrm{max}}\simeq4\mu$Hz and obtain $\log(g)$ estimates with a typical uncertainty below 0.05 dex, which is superior to typical measurements from spectroscopy. Additionally, the $\nu_{\mathrm{max}}$ distribution of our detections show good agreement with results from a simulated model of the Milky Way, with a ratio of observed to predicted stars of 0.992 for stars with 10$\mu$Hz $ <\nu_{\mathrm{max}}<270\mu$Hz. Among our red giant detections, we find 909 to be dwarf/subgiant stars whose flux signal is polluted by a neighbouring giant as a result of using larger photometric apertures than those used by the NASA \textit{Kepler} Science Processing Pipeline. We further find that only 293 of the polluting giants are known \textit{Kepler} targets. The remainder comprises over 600 newly identified oscillating red giants, with many expected to belong to the galactic halo, serendipitously falling within the \textit{Kepler} pixel files of targeted stars.

\end{abstract}




\section{Introduction}
Red giants showing solar-like oscillations from NASA's \textit{Kepler} mission \citep{Borucki2010} are critical to our understanding of stellar evolution and stellar populations. The asteroseismic study of such oscillations provides us with the ability to probe the stellar interior conditions such as core rotation rates \citep{Beck_2011,Mosser_2012,Deheuvels_2014}, and possibly core magnetic fields \citep{Fuller_2015, Stello_2016a,Stello_2016}, whose existence remains highly investigated (e.g. \citealt{Mosser_2017}). Moreover, the large number of oscillating red giants from \textit{Kepler} has provided us with the opportunity to study and characterize large stellar populations through ensemble analyses \citep{Chaplin_2013, Hekker_2017, Garcia_2018} that can inform galactic archaeology studies (e.g. \citealt{Casagrande_2015, Sharma_2016, Aguirre_2018}). 

From the 197,096 stars observed by \textit{Kepler} in long-cadence ($\Delta T \simeq 30$min.) from the \textit{Kepler} Data Release 25 (DR25), there is a current combined total of $\sim$19000 identified oscillating red giants reported in literature \citep{Yu_2018}. This sample comes from ensemble studies of previous data releases \citep{Hekker_2011, Huber_2011, Stello_2013} and newly identified targets \citep{Huber_2014, Mathur_2016, Yu_2016}, from which only $\sim$16000 have been analysed seismically based on the full end-of-mission 4-year light curves \citep{Yu_2018}. This number represents a significant fraction of the total number of \textit{Kepler} targets. However, recent population estimates from Gaia-derived radii on $\sim$178,000 \textit{Kepler} DR25 stars showed that the number of \textit{Kepler} red giants is approximately 21,000 \citep[hereafter B18]{Berger_2018}, proving that there is still a significant number of oscillating giants yet to be found from the end-of-mission data.

While red giants can be identified from their stellar parameters such as effective temperature, $T_{\mathrm{eff}}$, or surface gravity, $\log (g)$, the only way to guarantee that a giant shows solar-like oscillations is to directly detect the oscillations within its power spectrum, where they appear in the form of a Gaussian-like power excess on top of a sloping granulation background profile. Thus, here we perform a search for such oscillations across all long-cadence \textit{Kepler} targets from the DR25, a task that has not yet been done prior to this study.
The conventional way of detecting oscillations is carried out using seismic pipelines, where model fitting and statistical tests are performed on the power spectrum of the star (see \citealt{Stello_2017} for a description of pipelines to date). While rigorous, these approaches can experience difficulties in capturing the full complexity of the power spectra of real data, which are otherwise easily identified visually by a human expert. 

Instead, we use a non-conventional way of detecting oscillations using machine learning to aid our search for oscillating giants. In particular, we use the method introduced by \citet[hereafter H18]{Hon_2018}, which uses deep learning, an approach to artificial intelligence, to mimic human experts in visually identifying solar-like oscillations from plots of power density spectra. For brevity, from here onwards we use the term `power spectrum' to simply refer to the power density spectrum of the star. Besides identifying oscillating red giants, the frequency at maximum power of the star, $\nu_{\mathrm{max}}$, can also be measured by the deep learning visual expert based on the position of the oscillations within the power spectra. As shown by H18, this method achieves a high detection accuracy and is capable of predicting $\nu_{\mathrm{max}}$ with a human-level performance ($\sigma_{\nu_{\mathrm{max}}}/\nu_{\mathrm{max}} \sim 5\%$), while also being highly robust because no explicit model fitting of the power spectrum is required.

Our study in this paper is hence focused on the outcome of performing our classification over all $\sim$197,000 \textit{Kepler} targets ever observed during the mission, and characterizing the oscillating stars with the seismic parameter $\nu_{\mathrm{max}}$, which we then use to obtain $\log(g)$ estimates. A novel approach in our study includes the use of a larger photometric aperture, which produces an additional yield of serendipitous red giants. To date, this work represents the largest wide scale search for solar-like oscillations for giants with $\nu_{\mathrm{max}}\apprge4\mu$Hz, which will be an essential input to the full \textit{Kepler} legacy catalog for red giants (Garcia et al. in preparation).

\section{Methods}
First, we describe the preparation of the data and our specific use of a different photometric aperture, followed by the detection of oscillating stars within the \textit{Kepler} long-cadence dataset using deep learning methods.

\subsection{\textit{Kepler} Observations and Data Preparation}
\label{sec:data}
Our dataset comprises a total of 196,581 \textit{Kepler} end-of-mission long-cadence light curves. The remaining 515 stars from the DR25 were only observed for very short durations, thus they are not included in our study. Compared to the simple aperture photometry used by the \textit{Kepler} Science Processing Pipeline \citep{Jenkins_2010} that maximizes the signal-to-noise ratio for a specific \textit{Kepler} target within a 6.5 hour window \citep{Bryson_2010}, we use typically larger custom apertures that are designed to produce stable light curves on longer time scales \citep{Garcia_2014a}. More specifically, in small, noise-optimized apertures, the movement of stars across the spacecraft CCD can cause their light curves to have flux variations during a quarter and inter-quarterly flux discontinuities. This can significantly affect the oscillation frequency spectra of red giants. Using larger apertures allows us to mitigate this.

We define the extent of the aperture by moving outwards pixel-by-pixel from the center of the point-spread function of the target star and calculating a reference flux value of each subsequent pixel. To do this, we first convert flux values from units of electrons/second into dimensionless values by dividing with their uncertainties. We then compute the reference flux in a pixel as the 99.9th percentile flux value  of the pixel's light curve to avoid taking outliers into account.
If the reference flux value of a pixel is smaller than that of the previous pixel and is above a pre-defined threshold of 100, it is included in the aperture. However, if the reference flux is below the threshold, it is considered to contain only background signal. We then do not include the pixel in the aperture and stop extending the aperture in the direction of this pixel. A pixel is also excluded if its reference flux value is greater than the previous pixel because this indicates that there is another object in the vicinity of the target star. On average, our custom apertures have three times more pixels compared to the apertures used by the \textit{Kepler} Science Processing Pipeline. The amount of added noise is equal to the average noise from pixels containing low signal-to-noise ratios (SNR), and is dependent on the brightness of the target and the number of new pixels added in the aperture. However, we find that this amount of extra noise is small compared to the noise levels of \textit{Kepler} red giant targets from \textit{Kepler} Science Processing Pipeline data.

Once we have defined the extent of the aperture, we calculate the total flux within it for each cadence to generate the light curve. We treat each quarter separately to allow the apertures to change between spacecraft rolls. Finally, we correct the measured light curves by applying the methods described by \citet{Garcia_2011}. In particular, the light curves are high-pass filtered with a triangular filter with a cut-off providing a full transmission for signals shorter than 20 days and a smooth transition of the filter up to 40 days. No signals longer than this period remain in the data.


To increase the effectiveness of detecting solar-like oscillations, we need to prevent spectral leakage of high power from low frequencies, which is an effect of the spectral window that can obscure the intrinsic spectral frequency profile of convective granulation, oscillations, and white noise \citep{Garcia_2014b}. Hence, we use the in-painting technique based on a multi-scale cosine transform described by \citet{Pires_2015} to fill small gaps in the time series, which are mostly caused by regular missing points at a three day cadence from the angular momentum dump of the reaction wheels used for the attitude control of the spacecraft. While the gaps from the angular momentum dump can also be filled using linear interpolation over them, we find that the in-painting technique better reconstructs other gaps that can span several hours or a few days, which arise from instrumental effects and the monthly down-link of data from the \textit{Kepler} spacecraft \citep{Garcia_2014b}.

Once we obtain the power spectrum of a star, we convert it into a 2D grayscale image with a pixel dimension of 128x128 following the methods described by H18. The image is a white plot of the power  spectrum in logarithmic axes over a black background within the frequency range of $3\mu$Hz $ \leq \nu \leq 283\mu$Hz and a power density ($P$) range of $3$ppm$^2\mu$Hz $^{-1} \leq P \leq3 \times 10^7$ppm$^2\mu$Hz $^{-1}$.

\subsection{Deep Learning Methods}
\begin{figure*}
\centering
	\includegraphics[width=\linewidth]{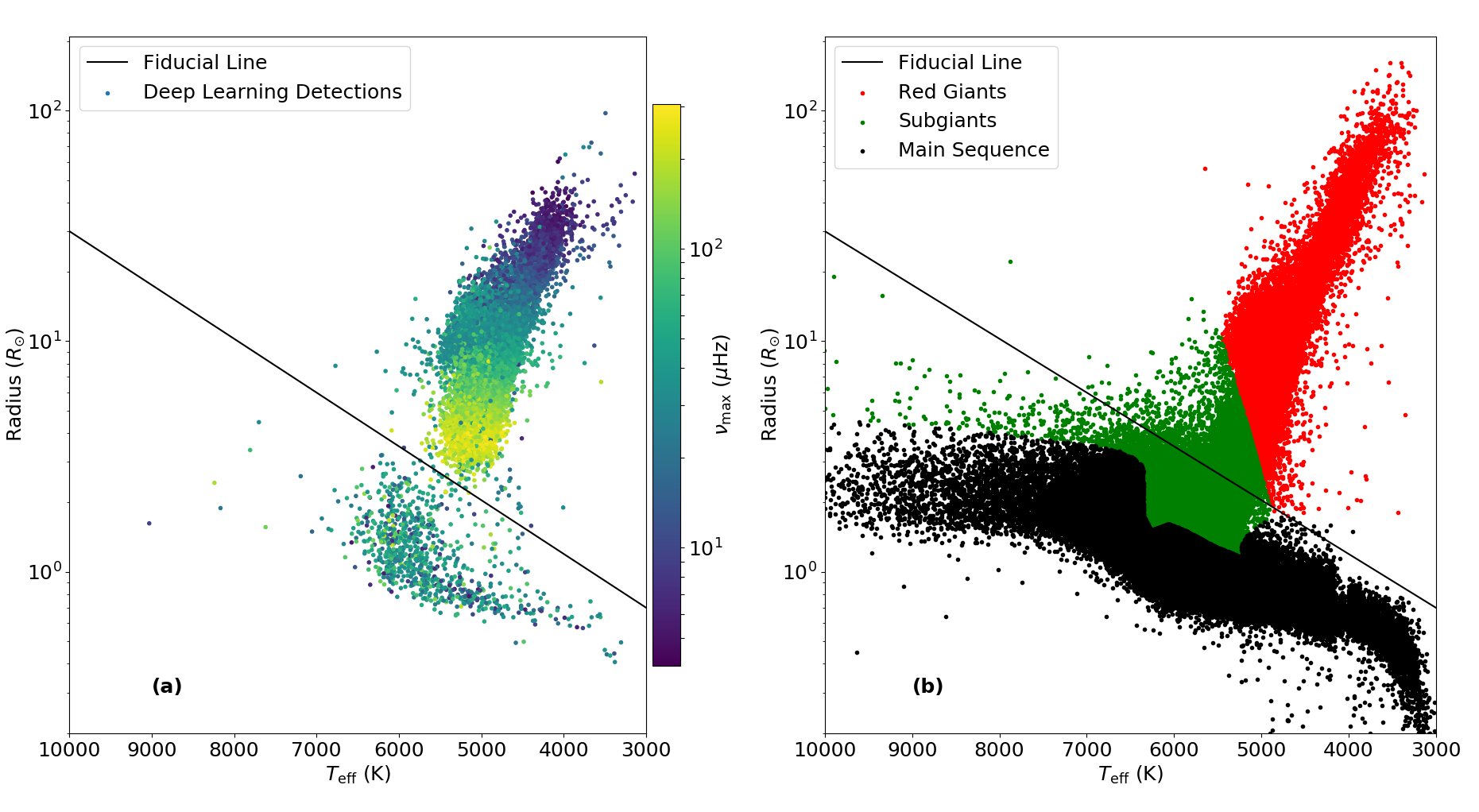}
	\caption{$R$-$T_\mathrm{eff}$ diagrams of \textit{Kepler} long-cadence targets, with values of $R$ and $T_\mathrm{eff}$ adopted from the catalog by B18. (a) Positive detections from our work, with points coloured by their measured $\nu_{\mathrm{max}}$ from the deep learning regressor. A fiducial line given by Equation \ref{fiducial_line} is drawn to delineate the boundary between the red giants and the non-red giants. 1260 detections do not have $R$ and $T_\mathrm{eff}$ listed in the catalog and hence are not plotted. (b) 177,911 \textit{Kepler} long-cadence targets from B18, coloured by their given evolutionary state from the same source.}

	\label{pred_gaia_rad}
\end{figure*}
\label{Detection_Method}

We use the 2D deep learning method as described by H18 to automatically detect the presence of red giant solar-like oscillations in the power spectrum of each star in the \textit{Kepler} long-cadence dataset. This method uses a combination of three 2D convolutional neural networks to detect specific image features within the 2D grayscale image of the power spectrum in order to classify it as either a positive detection (showing red giant solar-like oscillations) or a non-detection. In particular, the image of the power spectrum is required to show a broad hump of power excess on a granulation background to be classified as a positive detection by the classifier. 

The training set used to train the classifier is the same as that used by H18, comprising 31,123 \textit{Kepler} targets, with 15,924 known oscillating red giants analysed by \citet{Yu_2018} and the remaining 15,199 not showing red giant solar-like oscillations. However, their training set used light curves split into 3-month segments whereas our current training set uses full-length data. The classifier provides a probability, $p$, that a star shows oscillations, along with an uncertainty on the probability, $\sigma_p$, which is calculated using Monte Carlo methods. We determine $p =0.5$ as a suitable threshold above which a star is considered as a positive detection (see Section 2.4.1 in H18 for details). Once we determine which stars show oscillations, we predict their frequency at maximum power, $\nu_{\mathrm{max}}$, using the 2D deep learning regressor described by H18. Because the regressor learns using the same 2D input as the deep learning classifier, it generally uses the same contextual clues from the power spectrum as the classifier to give a visual estimate of the oscillation power excess location. This visual estimate is the prediction of $\nu_{\mathrm{max}}$ for the star, with a typical measurement uncertainty of 5\% as shown by H18 on red giants observed by the K2 mission \citep{Howell_2014}.  

Our deep learning classifier and regressor are constructed with the Keras library \citep{Keras} using the Tensorflow back end \citep{Tensorflow}, with each utilizing an NVIDIA Titan Xp GPU and the NVIDIA cuDNN library \citep{Chetlur_2014} for training and prediction. 

\subsection{Limitations of Current Classifier Version}\label{classifier_limitation}
We use a version of the deep learning classifier that is similar to that used by H18, which has been shown to be capable of achieving a detection accuracy of 98-99\% when classifying red giants observed during the K2 mission. However, the task of predicting on the entire \textit{Kepler} long cadence dataset imposes several difficulties towards generalizing this level of performance towards all \textit{Kepler} targets.

\subsubsection{Varied Observation Lengths}\label{observation_length}
Because not all long-cadence \textit{Kepler} targets were observed across 4 years, the current classifier is likely to predict stars observed for fewer than two quarters as non-detections due to their low frequency resolution. This would therefore include genuine oscillating giants, which we denote as false negatives. To reduce false negatives among these shortly-observed stars, we use a version of the classifier that is trained on 3-month segments of \textit{Kepler} light curves to run a second pass over them. Since we use a separate classifier selectively, we cannot assume that the combined use of 3-month and 4-year classifiers will obtain exactly the same level of accuracy as reported for the classification of K2 giants.

\subsubsection{Differences in Dataset Complexity}\label{Dataset_Complexity}
Our current version of the deep learning classifier has been conditioned to perform well on K2 targets. However, compared to K2, \textit{Kepler} targets are more biased towards hot dwarfs compared to K2 targets \citep{Huber_2016}, resulting in a larger fraction of stars near the instability strip. Thus, we expect a higher number of variable stars including $\delta$ Scuti pulsators or hybrid variables in the full \textit{Kepler} dataset. This results in a wider range of complex frequency-power profiles that has to be identified by the classifier, causing it to potentially make mistakes on examples it has rarely seen during training. For instance, it may incorrectly identify a star that does not show giant oscillations as a positive detection, which we label as a false positive. Although our current classifier has trained on a large range of variable star types, it is possible that certain types within the full \textit{Kepler} long cadence data are still under-represented.
Another type of incorrect prediction potentially made by the classifier is a false negative, where it classifies a genuine oscillator as a non-detection. This can happen if there is a superposition of the oscillation power excess with signatures from other forms of variability that show multiple sharp peaks in the power spectrum, such as binarity or classical pulsations.


One approach to resolve this problem would be to retrain the network with a training set that includes many more classical pulsators, resulting in more examples with complex power spectra. However, such a modification to the network would require additional considerations with respect to the quantity and the variety of stellar types along with potential changes to the architecture of the neural network, which is beyond the scope of this paper. A simpler and quicker approach that we use in this study is to first perform an initial detection with the classifier, followed by visual inspection of certain suspicious predictions in order to refine the list of detections. We discuss the identification of these suspicious predictions in Section \ref{refine_detections}. Using this approach, the role of the classifier in this study is to significantly narrow down the search space for the identification of oscillating giants, resulting in only a small subset of predictions that have to be confirmed by visual inspection. 

\section{Detection of Oscillating Red Giants}
In this Section, we obtain the results from the detection process, obtain $\log (g)$ measurements for each detection, analyze the $Kp-\nu_{\mathrm{max}}$ distributions of the detected oscillating giants, and determine the completeness of our oscillating red giant sample.


\subsection{Obtaining a Final List of Detections}\label{refine_detections}
\subsubsection{Preliminary Detections from Deep Learning Classifier}
We initially detect solar-like oscillations in 21,468 stars with the deep learning classifier.
To account for stars observed for shorter durations (Section \ref{observation_length}), we use the 3-month classifier to identify 1,130 false negative predictions made by the 4-year classifier. As a result, we identify a total of 22,598 detections from deep learning methods. 

\subsubsection{Identification of Suspicious Predictions}
In order to identify potentially inaccurate predictions caused by differences in dataset complexity (Section \ref{Dataset_Complexity}), we cross-match our list of detections with stars from the \textit{Kepler} Eclipsing Binary Catalog \citep{Kirk_2016}\footnote{\href{http://keplerebs.villanova.edu/}{http://keplerebs.villanova.edu/}} and filter out 90 detached binaries that are false positives after visual inspection of their power spectra. Next, we plot a $R-T_\mathrm{eff}$ diagram for our detections using values from the Gaia (DR2)-\textit{Kepler} (DR25) cross-matching catalog by B18, shown in Figure \ref{pred_gaia_rad}a. We find two well-separated groups of stars with a boundary that we delineate with a fiducial line, given by:
\begin{equation}\label{fiducial_line}
R = \bigg(\frac{T_\mathrm{eff}-3000}{7000}\bigg) \log\bigg(\frac{300}{7}\bigg) \times 10^6 R_{\odot}.
\end{equation}

As seen in Figure \ref{pred_gaia_rad}b, detections above the line evidently correspond to red giants with $R \apprge 2R_{\odot}$ and $T_\mathrm{eff} \apprle 6250$ K, while detections below the line correspond to subgiants or main sequence stars that have $R \apprle 2R_{\odot}$. We note that our fiducial line coincides with the minimum radius for a red giant following the B18 classification. While we initially detect 1420 stars below the fiducial line, we filter out 550 false positives (from which many are detached eclipsing binaries and $\delta$ Scutis) by visual inspection. The remaining 870 stars show red giant solar-like oscillations but have radii representative of main sequence/subgiant stars.

\subsubsection{Validation by Seismic Pipeline}\label{pipeline_validation}
To further ensure the validity of our results, we cross-match our results with a list of subgiant and red giant detections from the A2Z seismic pipeline \citep{Mathur_2010}. The detections from A2Z are based on the same light curves used for the deep learning classification. A2Z performs a detection of excess power based on the measurement of the mean large separation of oscillation modes and the position of the maximum power in the power spectrum. A detection is confirmed when these two measurements follow the $\Delta\nu$-$\nu_{\mathrm{max}}$ relation \citep{Stello_2009} to within 10\%. A2Z also uses the FliPer metric \citep{Bugnet_2018} to find outliers on which the pipeline is then re-run to search for modes in the frequency range predicted by FliPer. 

We find 3847 stars that are predicted as detections by A2Z but not by the deep learning classifier. From these, we ignore the 2444 stars with $\nu_{\mathrm{max}} \apprle 3\mu$Hz and 1210 stars with $\nu_{\mathrm{max}} > 283\mu$Hz because these are not detectable by our current version of the deep learning classifier by construction. By visual inspection of the power spectra of the remaining 193 stars, we confirm that the deep learning classifier missed these genuine oscillators (false negatives). As described in Section \ref{Dataset_Complexity}, most of these have detached binary signals or classical pulsations superimposed over their oscillation power excess. We find 39 of these genuine oscillating giants have Gaia-radii representative of main sequence/subgiant stars.

We also find 1195 stars predicted as detections by the deep learning classifier but not by A2Z. Again, we visually inspect their power spectra and find 135 false positives by the deep learning classifier. Thus, the remaining 1060 stars are genuine oscillating red giants detected by our deep learning classifier that have not been detected by A2Z.

\subsubsection{Final Tally of Oscillating Giants}

After adding false negatives and removing false positives from our list of detections, we have a total of 21,914 oscillating red giants, each with a measured $\nu_{\mathrm{max}}$ from the deep learning regressor. From these, 21,721 are detected by the deep learning classifier and hence are given a detection probability, $p$, and its associated uncertainty, $\sigma_p$. The remaining 193 giants are false negatives that have been manually included as described in Section \ref{pipeline_validation}. For these we are, however, still able to provide a $\nu_{\mathrm{max}}$ from the deep learning regressor. Using the probability distribution $p \pm \sigma_p$, we define stars as marginal detections if they have an average likelihood above 33\% of having $p<0.5$ (in other words, they are classified as non-detections more than 1/3 of the time). This results in 809 marginal detections from the deep learning classifier.

20,654 of our detected oscillating giants have $T_{\mathrm{eff}}$ and $R$ from the B18 catalog and are plotted in Figure \ref{pred_gaia_rad}a, where it is shown that we can detect oscillations in red giants up to $\sim40R_{\odot}$ (see discussion in Section \ref{distributions}). We determine a total of 909 main sequence/subgiant stars showing red giant solar-like oscillations, which we will address in Section \ref{Stars_Below_Fiducial_Line}. We also find 1,671 stars with $R\leq40R_\odot$ that are classified as red giants by B18 but are not in our list of detections. These stars are of interest because they may be experiencing suppression of their oscillations. We present these stars in Appendix \ref{RG-NonDet}.

Our results are tabulated in Table \ref{Detection_Table} (available online). Besides the outputs from the deep learning methods ($p, \sigma_p, \nu_{\mathrm{max}}$), we also indicate marginal detections, and whether the star is above the fiducial line (has red giant Gaia-derived radii) or below it (has dwarf/subgiant Gaia-derived radii). The Table also includes the estimated surface gravity, $\log (g)$, for each detection, which we discuss in Section \ref{estimating_surface_gravity}. 

\begin{table} 
			\centering
			\caption{\textit{Kepler} long-cadence stars showing red giant solar-like oscillations, identified by their \textit{Kepler} Input Catalog (KIC) ID \citep{KIC}. Values in brackets for $\nu_{\mathrm{max}}$ and $\log (g)$ are 1$\sigma$ uncertainties. The flag `0' indicates stars above the fiducial dwarf/giant-separating line in Figure \ref{pred_gaia_rad}, flag `1' indicates stars below the fiducial line, flag `2' indicates stars that do not have $T_{\mathrm{eff}}$ and $R$ from the B18 catalog, while the additional flag `M' indicates a marginal detection. The value `-99' is assigned to $p\pm\sigma_p$ for false positives that have been rectified by visual verification, and to $\log(g)$ for stars that have no measured $T_{\mathrm{eff}}$ from B18 or from the \textit{Kepler} DR25 Stellar Properties Catalog \citep{Mathur_2017}. The full version of this table is available in a machine-readable form in the online journal, with a portion shown here for guidance regarding its form and content.
			}
			\label{Detection_Table}
			\begin{threeparttable}
				\begin{tabular}{|c|c|c|c|c|c|}
					\hline
					KIC ID &$p$&$\sigma_p$&$\nu_{\mathrm{max}} (\mu$Hz)&$\log(g)$ (dex)&Flag\\
					\hline
					5611229&1.000&0.000&54.44 (1.53)&2.66 (0.01)&0\\
					5611572&0.667&0.445&58.65 (5.25)&2.71 (0.03)&1M\\
					5616489&1.000&0.000&30.44 (1.30)&2.40 (0.02)&0\\
					5616491&0.926&0.199&36.83 (7.06)&2.53 (0.09)&1\\
					5616606&1.000&0.000&35.33 (1.20)&2.47 (0.02)&0\\
					5621709&0.892&0.221&40.59 (4.14)&2.57 (0.04)&1\\
					5629090&0.977&0.126&7.01 (0.36)&1.73 (0.02)&0\\
                    5613033&0.999&0.002&80.6 (2.83)&2.87 (0.02)&2\\
					5648159&1.000&0.000&205.10 (6.34)&3.23 (0.01)&0\\
                    5648794&-99&-99&33.01 (1.34)&-99&2\\
					...&...&...&...&...&...\\
					\hline
					
				\end{tabular}
			\end{threeparttable}
		\end{table}

\subsection{Estimating the Surface Gravity of Detected Stars}\label{estimating_surface_gravity}
\begin{figure}
	\centering
	\includegraphics[width=1.1\linewidth]{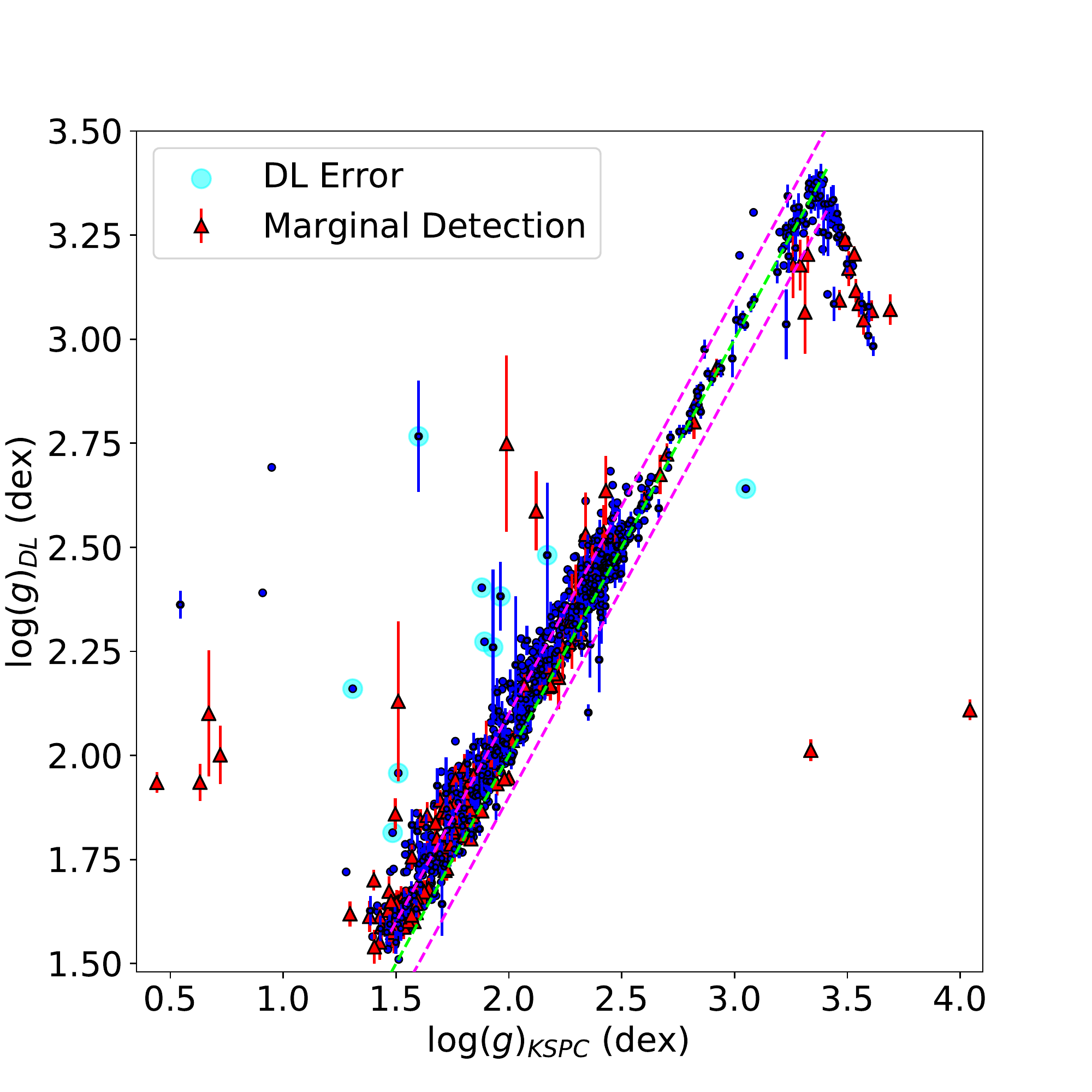}
	\caption{A comparison of estimated surface gravity values from the deep learning regressor, $\log(g)_{\mathrm{DL}}$, with asteroseismic surface gravity measurements from the \textit{Kepler} DR25 Stellar Properties Catalog \citep{Mathur_2017}, $\log(g)_{\mathrm{KSPC}}$, for the 1430 stars in our validation sample. The green dashed line indicates a one-to-one relation, while the magenta dashed lines indicate $\pm 0.1$ dex deviations from the green line. For clarity, the errorbars of only every 5th blue point is plotted. Marginal detections are plotted in red, while predictions with large deviations from the one-to-one relation ($>0.25$ dex) that are visually verified to have inaccurately predicted $\nu_{\mathrm{max}}$ are highlighted in cyan.}
	\label{log_g_plot}
\end{figure}

\begin{figure*}
\centering
	\includegraphics[width=0.95\linewidth]{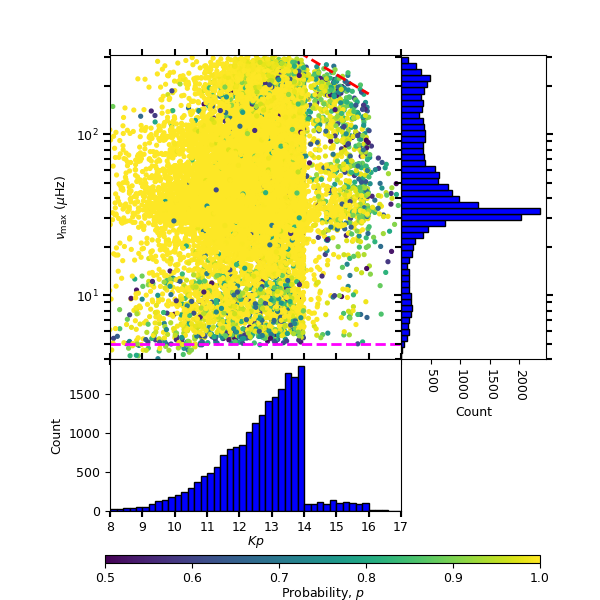}
	\caption{Measured $\nu_{\mathrm{max}}$ vs. $Kp$ for detections above the fiducial line in Figure \ref{pred_gaia_rad}. The colours of each point indicate the probability of a detection, $p$, as given by the deep learning classifier. The distributions of $Kp$ and $\nu_{\mathrm{max}}$ are shown as histograms in the bottom and right panels, respectively. The dashed red line indicates an empirical detection limit $\nu_{\mathrm{max-lim}} < 1.5\times 10^4 \cdot 2^{-0.4\cdot Kp}$, above which the signal-to-noise is too low for the detection of solar-like oscillations.The apparent discontinuity in the detection limit at \textit{Kp} = 16 down to $\nu_{\mathrm{max}}=70\mu$Hz is most likely related to target selection rather than an actual detection bias (see text). The magenta dashed line at $\nu_{\mathrm{max}}=5\mu$Hz indicates the minimum $\nu_{\mathrm{max}}$ of the training set used to train the deep learning classifier and regressor, which imposes a limitation on the range of our measurements.}

	\label{posdet_nonblend_hist}
\end{figure*}

From the measured $\nu_{\mathrm{max}}$ of each oscillating giant, we compute its surface gravity, $\log(g)$, using the following scaling relation \citep{Brown_1991,Kjeldsen_1995}:

\begin{equation}
\frac{g}{g_{\odot}} \simeq \bigg( \frac{\nu_{\mathrm{max}}}{\nu_{\mathrm{max\odot}}} \bigg)\bigg( \frac{T_{\mathrm{eff}}}{T_{\mathrm{eff\odot}}} \bigg)^{1/2},
\label{scaling_relation}
\end{equation}

\noindent where we use $\log(g_{\odot}) = $ 4.44 dex, $\nu_{\mathrm{max\odot}} = 3050$ $\mu$Hz and $T_{\mathrm{eff\odot}} = 5772$K. We adopt $T_{\mathrm{eff}}$ and $\sigma_{T_{\mathrm{eff}}}$ values from the B18 catalog if available, otherwise we use values from the \textit{Kepler} DR25 stellar properties catalog \citep[hereafter KSPC]{Mathur_2017}. We include our calculated $\log(g)$ and the associated uncertainties in Table \ref{Detection_Table}.

To validate our estimated $\log(g)$ values, we compare them with those from the KPSC. We only select stars that have measured $\log(g)$ from asteroseismology from the KSPC because such measurements are the most precise. Additionally, we filter out stars that are also in our training set to prevent a biased validation result. As a result, our validation sample comprises 1430 stars with asteroseismic $\log(g)$ values that the deep learning regressor has not `seen' during training. 

The comparison of our $\log(g)$ values, $\log(g)_{\mathrm{DL}}$, with those from the validation sample, $\log(g)_{\mathrm{KSPC}}$, is shown in Figure \ref{log_g_plot}. In general, we find that 88\% and 96\% of $\log(g)_{\mathrm{DL}}$ lie within $\pm 0.05$ dex and $\pm 0.1$ dex of $\log(g)_{\mathrm{KSPC}}$, respectively. For $2.0$ dex $\leq \log(g)_{\mathrm{DL}} \leq 2.6$ dex, we find $\log(g)_{\mathrm{DL}}$ to be overestimated by $\sim0.04$ dex, with $\sim$75\% of stars within 2$\sigma$ of $\log(g)_{\mathrm{KSPC}}$. For $\log(g)_{\mathrm{DL}} < 2.0$ dex, this overestimation increases to $\sim0.08$ dex, with $\sim$50\% of stars within 2$\sigma$ of $\log(g)_{\mathrm{KSPC}}$. 

For stars with $\log(g)_{\mathrm{DL}} \apprle 2.8$ dex, we visually inspect the power spectra of all 25 stars with unusually large deviations (> 0.25 dex) from the one-to-one relation (green line). We find that 10 of these  have inaccurate $\nu_{\mathrm{max}}$ measurements from the deep learning regressor (highlighted in cyan), while 9 stars have very low signal-to-noise levels and have been flagged marginal detections (red points). For the remaining 6 stars, we find that our measured $\nu_{\mathrm{max}}$ are consistent with the observed frequency range of the oscillation power excess within 2$\sigma$. 

At $\log(g)_{\mathrm{DL}} > 2.8$ dex, we observe a downward `trail' of 30 stars with large $\log(g)$ deviations. This feature appears to approximate a reflection of the one-to-one relation around $\log(g)_{\mathrm{KSPC}}\simeq3.5$ dex, indicating that the detected oscillation power excess for these stars are potentially aliased counterparts of oscillations with frequencies beyond the Nyquist frequency of $\sim$283$\mu$Hz \citep{Yu_2016}. This results in a smaller $\log(g)_{\mathrm{DL}}$ compared to $\log(g)_{\mathrm{KSPC}}$. While 13 of these have been flagged as marginal detections, we verify visually that the remaining have $\nu_{\mathrm{max}}$ values consistent with the frequency range of the oscillation power excess in their power spectra.

In Figure \ref{log_g_plot}, we find more marginal detections for stars with $\log(g)_{\mathrm{DL}} \apprle 2.0$ dex compared to the rest of our validation sample. Those with very low surface gravities ($\log(g)_{\mathrm{DL}} \apprle 1.75$ dex) correspond to stars with $\nu_{\mathrm{max}}$ values close to the low-frequency detection limit of the classifier at $\sim4\mu$Hz. At this limit, the classifier is less confident in identifying solar-like oscillations because they appear just within the frequency range of the 2D image of the power spectra. Meanwhile, we find that many marginal detections with $1.75$ dex $\apprle \log(g)_{\mathrm{DL}} \apprle 2.0$ dex ($6\mu$Hz $\apprle \nu_{\mathrm{max}} \apprle 15\mu$Hz) correspond to the stars that were observed for very short durations (typically Quarters 0 and 1) but have power spectra that show large granulation power with a power-frequency profile resembling solar-like oscillations despite their low frequency resolution. Notably, while these stars have been flagged as marginal detections, most have $\log(g)_{\mathrm{DL}}$ values within 0.2 dex of $\log(g)_{\mathrm{KSPC}}$ and thus are not considered major outliers. Moreover, because every star in the validation set has been confirmed by the KSPC to show solar-like oscillations, this indicates that while the classifier's predictions for low $\nu_{\mathrm{max}}$ stars tend to be uncertain, many of them tend to be genuine oscillating giants.

In summary, to select stars with reasonably accurate surface gravity estimates with respect to the KSPC, we recommend using the uncertainties from the deep learning regressor, $\sigma_{\log(g),\mathrm{DL}}$. In general, $\sim$96\% of our non-outlier measurements (blue) have $\sigma_{\log(g),\mathrm{DL}} \leq 0.05$ dex, while inaccurate $\log(g)_{\mathrm{DL}}$ values have $\sigma_{\log(g),\mathrm{DL}} > 0.05$ dex. Hence, selecting stars with $\sigma_{\log(g),\mathrm{DL}}$ below this threshold will generally provide reliable surface gravity estimates.

\subsection{Analysis of Oscillating Giants with Red Giant Gaia-Derived Radii}

\begin{figure*}
	\centering
	\includegraphics[width=\linewidth]{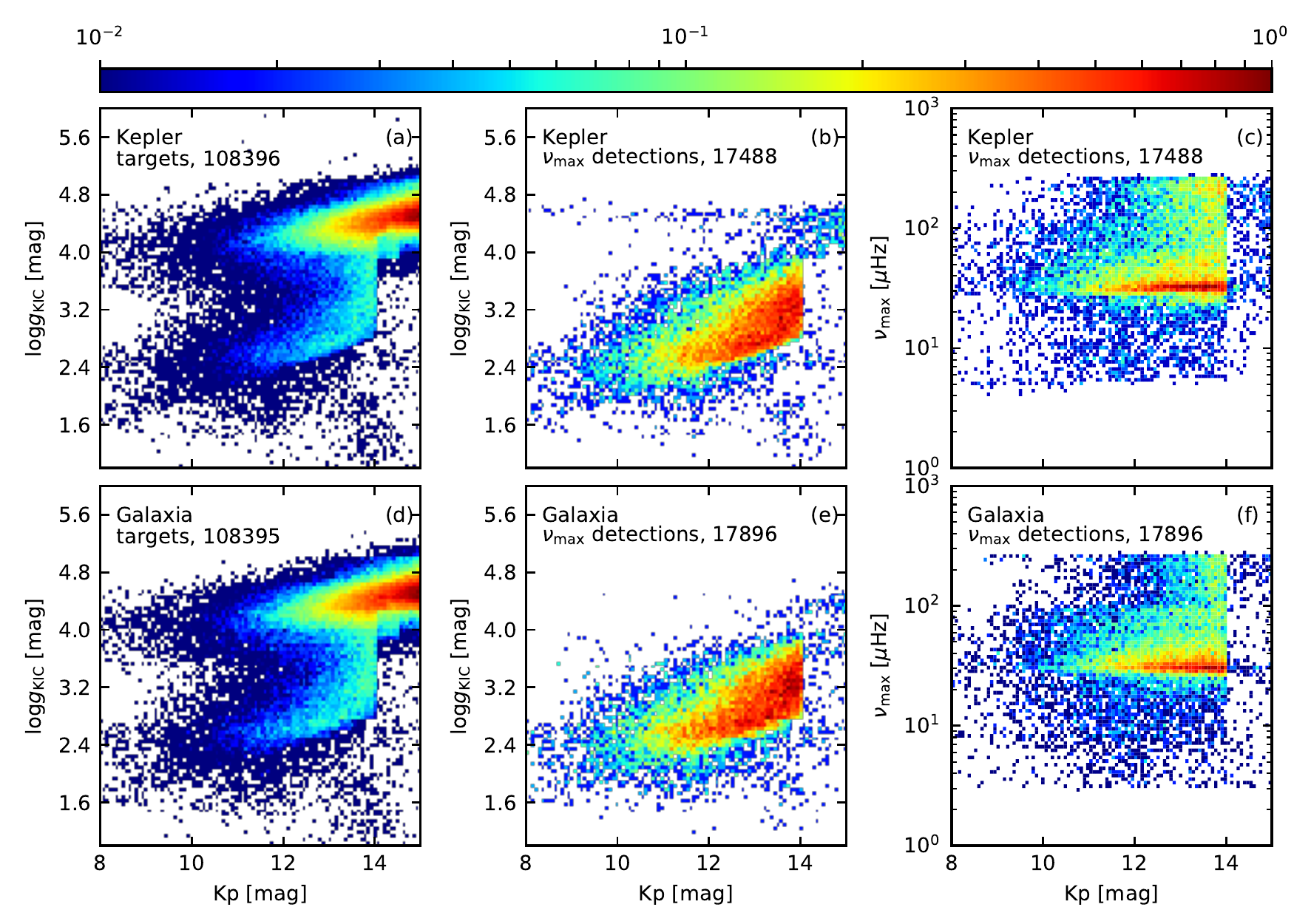}
	\caption{Comparison of the number of observed oscillating \textit{Kepler} giants from this study with predictions from Galaxia. Panels in the top row show results from \textit{Kepler} while panels in the bottom row show results for stars simulated with Galaxia that are sampled to match the distribution of \textit{Kepler} targets in $(\log g_{\rm KIC},Kp)$ space. Within each panel, the colour represents target density normalized by its maximum density value, while the listed number indicates the number of stars.  (a), (d): Distribution of all \textit{Kepler} targets in $(\log g_{\rm KIC},Kp)$ space. (b), (e): Distribution of targets with detected solar-like oscillations with $3\mu$Hz $<\nu_{\rm max}<270\mu$Hz in $(\log g_{\rm KIC},Kp)$ space. (c), (f): Same as panels (b) and (e), but in $(\nu_{\rm max},Kp)$ space.}
	\label{GalaxiaComp}
\end{figure*}

\subsubsection{Distributions of $\nu_{\mathrm{max}}$ and \textit{Kepler} magnitude} \label{distributions}
As expected, a comparison between the stars above the line in Figures \ref{pred_gaia_rad}a and \ref{pred_gaia_rad}b shows that our detections lack stars with $R \apprge 40R_{\odot}$. This is due to the inherent limitation of our method that is caused by our training set having no stars with $\nu_{\mathrm{max}} < 5\mu$Hz. Despite this, we note that our method can still detect stars down to $\nu_{\mathrm{max}} \simeq 4\mu$Hz, which is seen in the $\nu_{\mathrm{max}}$ vs. \textit{Kepler} magnitude ($Kp$) diagram in Figure \ref{posdet_nonblend_hist}. However, the classifier tends to be less confident at classifying stars near this frequency limit, seen as an increase in stars with $p<0.9$ (darker-coloured dots) directly above the $\nu_{\mathrm{max}} = 4\mu$Hz line in Figure \ref{posdet_nonblend_hist}. Aside from this low-frequency limit, we also see a significant number of stars with less confident (relatively low $p$) predictions at $Kp < 14$ up to $\nu_{\mathrm{max}}\simeq15\mu$Hz. These are consistent with our findings in Section \ref{estimating_surface_gravity}, where we found more marginal detections for stars with $\log(g)_{\mathrm{DL}} \apprle 2.0$ dex due to their short time series. Meanwhile, low-probability stars with $Kp < 14$ and 150$\mu$Hz $\apprle\nu_{\mathrm{max}}\apprle220\mu$Hz in Figure \ref{posdet_nonblend_hist} correspond to the stars that are suspected to be oscillating at super-Nyquist frequencies with $\log(g)_{\mathrm{DL}} > 2.8$ dex as shown in Figure \ref{log_g_plot}.

In Figure \ref{posdet_nonblend_hist}, we also see a decrease of the number of stars with $\nu_{\mathrm{max}} \apprge 230\mu$Hz and $Kp \geq 14$ (top right corner), which is due to the lower signal-to-noise levels with fainter magnitudes and higher $\nu_{\mathrm{max}}$ that limit the detectability of the oscillations. The lower signal-to-noise levels in this region of $\nu_{\mathrm{max}}-Kp$ space also result in fewer confident predictions made by the classifier. The red line in Figure \ref{posdet_nonblend_hist} shows an empirical approximation to the limit above which no detections can be found by our method. This limit appears to have a discontinuity (drop) at $Kp\simeq 16$, which is likely caused by stars at $Kp\apprge16$ originating from a different stellar sample that prioritized the observation of cool dwarfs (e.g. \citealt{Batalha_2010}). Hence, we infer that this discontinuity is from the selection effect of the sample rather than a detection bias/limit of our method. 

To make sure that the slope of this detection limit is not influenced by potentially spurious $\nu_{\mathrm{max}}$ measurements for low-luminosity stars, we investigate $\sim 20$ stars that lie just above the fiducial line in Figure \ref{pred_gaia_rad}a but have a measured $\nu_{\mathrm{max}} \apprle 200\mu$Hz.
We find that such stars are not predominantly faint and thus inaccurate $\nu_{\mathrm{max}}$ values do not influence the slope of the observed detection limit. 

Additionally, we note that this limit is consistently above the detection limit for the K2 mission, which varies as $2.6\times 10^6 \cdot 2^{-\cdot Kp}$ \citep{Stello_2017}. This is expected because K2 data generally have lower signal-to-noise levels compared to that from \textit{Kepler}.

\subsubsection{Detection completeness}
We estimate the detection completeness of the detected oscillating giants in our study 
by comparing with predictions from theoretical models of the Milky Way. For this purpose, we use Galaxia \citep{Galaxia} to simulate a mock \textit{Kepler} survey, from which we resample stars to match the selection function of the \textit{Kepler} targets, which is an approximate function that delineates all \textit{Kepler} giants in \textit{Kepler} Input Catalog (\citealt{KIC}, KIC) radius-\textit{r} magnitude space (see \citealt{Sharma_2016}, their Section 3.1). Here, we use the most recent version of Galaxia, which has improved and updated physical parameters (Sharma et al. in preparation). In particular, it has a thick disc 
with mean $\log Z/Z_{\odot}$ of -0.18 to match the
mean $\log Z/Z_{\odot}$ of stars between $1$ kpc $<|z|<2$ kpc and $5$ kpc $<R<7$ kpc in the GALAH survey \citep{Buder_2018}.

Because stars in the \textit{Kepler} mission were mainly selected based on \textit{Kepler} magnitude, \textit{Kp}, and photometric surface gravity, $\log g_{\rm KIC}$ \citep{Batalha_2010,Sharma_2016}, we assume the selection to be a function of $Kp$ and $\log g_{\rm KIC}$. To simulate this distribution of stars using Galaxia, we bin the full list of \textit{Kepler} targets into 28 x 48 bins in $(Kp,\log g_{\mathrm{KIC}})$ space as shown in Figure \ref{GalaxiaComp}a. Next, we resample stars from Galaxia to match the number of \textit{Kepler} targets as shown in Figure \ref{GalaxiaComp}d. Details of estimating $\log g_{\rm KIC}$ for Galaxia stars are described in \citet{Sharma_2016}. In short, the synthetic photometry in the $g$ band is re-calibrated to match the \textit{Kepler} pass bands, where the \textit{Kepler} Input Catalog Bayesian scheme \citep{KIC} is then used to estimate the stellar parameters. 

Next, we use the \citet{Chaplin_2011} method to estimate the probability of detecting $\nu_{\rm max}$ for a light curve spanning 4 years, $p_{\rm detect}$. We then compare the number of detected oscillating giants from \textit{Kepler} (Figure \ref{GalaxiaComp}b) with the number of Galaxia stars having $p_{\rm detect}>0.95$ (Figure \ref{GalaxiaComp}e). The $\nu_{\rm max}-Kp$ distribution for both observed and synthetic oscillating giants is shown in Figures \ref{GalaxiaComp}c and \ref{GalaxiaComp}f, respectively. From these results, we find that the distribution of \textit{Kepler} stars in 
$(Kp,\log g_{\mathrm{KIC}})$ and $(\nu_{\rm max},Kp)$ space matches well 
with predictions from Galaxia. Notably, a majority of \textit{Kepler} targets that show red giant oscillations in Figure \ref{GalaxiaComp}b have $\log g_{\mathrm{KIC}}<4$, which is in agreement with the distribution shown by Galaxia in Figure \ref{GalaxiaComp}e. However, in Figure \ref{GalaxiaComp}f we find that Galaxia predicts a sharp difference in the number of stars just below and above $\nu_{\rm max}\simeq 100 \mu{\rm Hz}$, which is not observed for the \textit{Kepler} targets in Figure \ref{GalaxiaComp}c. This is because the predicted population of secondary clump stars in Galaxia appear as an overdensity with a sharp maximum $\nu_{\mathrm{max}}$ cutoff. As the comparison shows, this cutoff is potentially too sharp to describe the population of observed secondary clump stars.

We find the ratio of observed to predicted stars to be 0.977 for stars with $3\mu{\rm Hz}<\nu_{\rm max}<270\mu{\rm Hz}$, and 0.992 for stars with $10\mu{\rm Hz}<\nu_{\rm max}<270\mu{\rm Hz}$. Hence, under the assumption that the Galaxia model provides a good representation of the true population, our sample of detected oscillating giants seems to be nearly complete except for stars with 3$\mu$Hz $<\nu_{\rm max}<10\mu{\rm Hz}$, which is expected due to the current limitation of our deep learning methods as discussed in Section \ref{distributions}.

\section{Dwarf/Subgiant Stars with Detected Giant Oscillations}\label{Stars_Below_Fiducial_Line}
In this Section, we address the 909 main sequence/subgiant stars in Section \ref{refine_detections} that show red giant oscillations. The following subsections detail how we identify these stars as blended targets.

\subsection{Blended Targets}
\label{determine_pollutants}
\begin{figure}
	\centering
	\includegraphics[width=\linewidth]{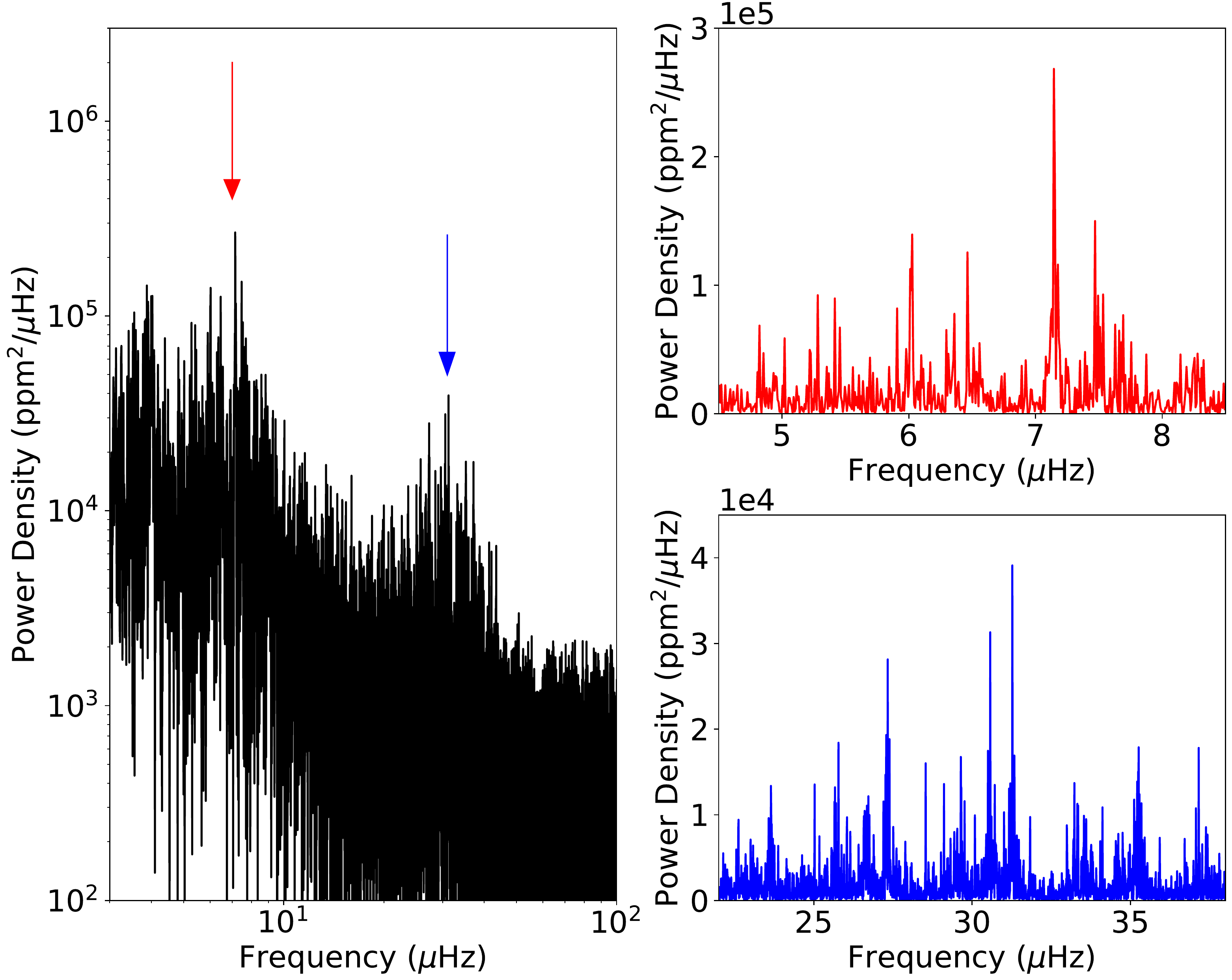}
	\caption{Power spectrum of a blended target, KIC 5824221. The two power excesses showing solar-like oscillations denoted by the arrows (left panel) are determined to be caused light pollution from nearby red giants KIC 5824237 (red) and KIC 5824232 (blue), respectively, with their corresponding oscillation modes in the spectrum shown in the right. The identification of these polluting giants is shown in Appendix \ref{Appendix_Double_Blend}.}
	\label{ExampleBlend}
\end{figure}
\begin{figure*}
\centering
	\includegraphics[width=\linewidth]{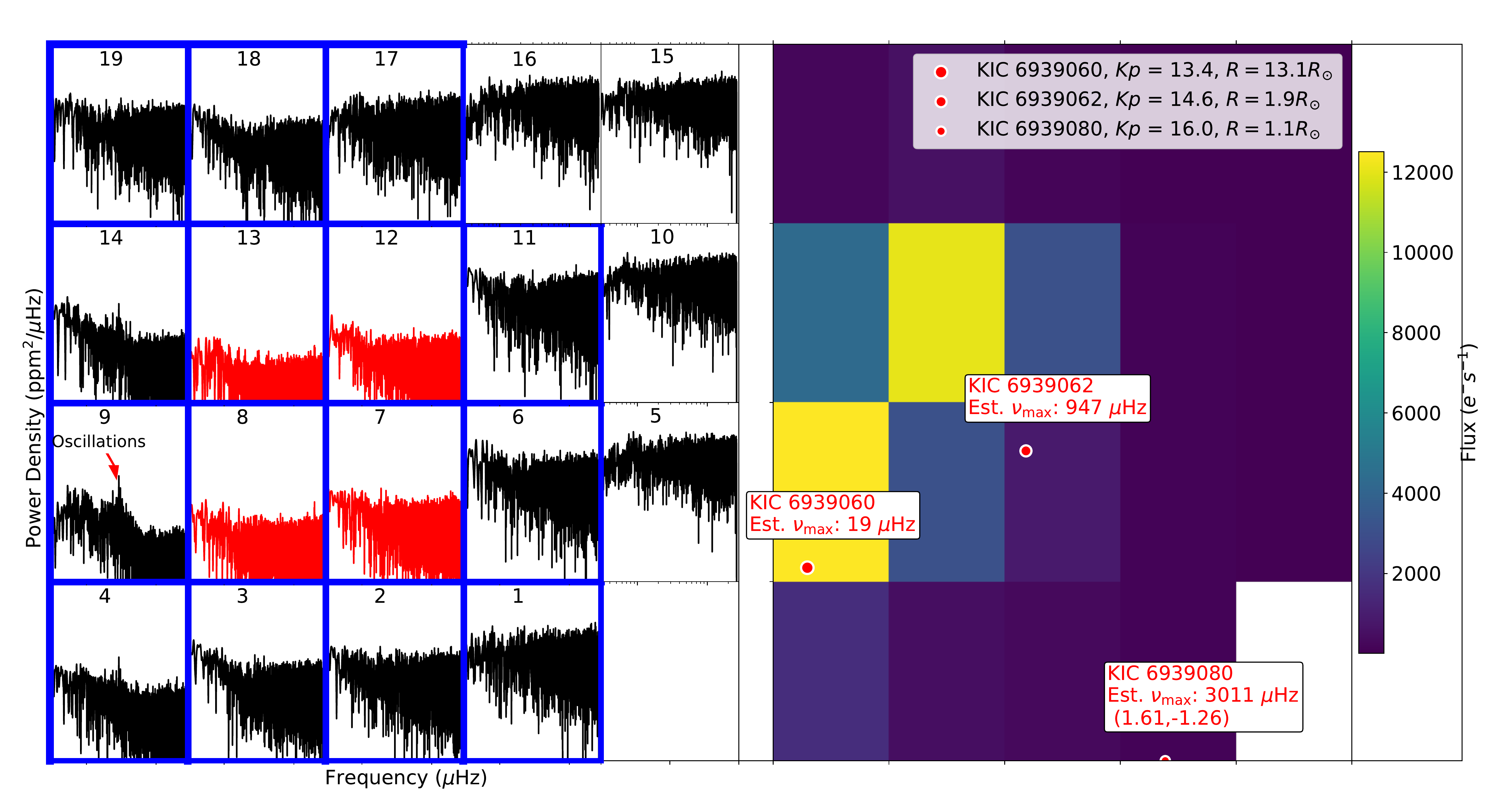}
	\caption{Target Pixel File (TPF) for Quarter 4 of the blended target KIC 6939062, with its power spectrum map (left) and flux map (right). A star map with the positions of the \textit{Kepler} target and nearby stars are overlaid on the flux map, with plot point sizes proportional to the negative logarithm of the \textit{Kepler} magnitude, $Kp$. The pixel aperture mask used in this study is highlighted with blue borders, while the mask used by the \textit{Kepler} Science Processing Pipeline is shown by red plots in the power spectrum map. KIC 6939060 is identified as the polluting star because it has a $\nu_{\mathrm{max,sc}} \simeq 19\mu$Hz using Equation \ref{scaling_relation}, which matches the frequency of the oscillation power excess that can be seen in pixels 4, 9, and 14 in the power spectrum map, indicated by the arrow in the left panel. These pixels also correspond to a high light intensity in the flux map, which are indicated by bright colours (right panel). Another star, KIC 6939080, is located beyond the bounds of the pixel image but is still within our search boundaries for \textit{Kepler} targets (5 pixels beyond TPF) and hence is still shown at the border of the star map.}

	\label{Maps_Known}
\end{figure*}

As a consequence of using a larger photometric aperture on the \textit{Kepler} targets, there is a greater likelihood of light pollution, where flux from a nearby star is captured by the aperture as well. Although this is mitigated when defining the extent of our aperture as described in Section \ref{sec:data}, flux from a nearby star may still be captured. As a result, the oscillation signal from this polluting star can be superimposed in the power spectrum of the original target as shown in Figure \ref{ExampleBlend}.

To confirm that the 909 dwarf/subgiant stars showing giant oscillations are indeed blended targets, we have to identify the polluting star for each target by searching for other nearby stars within its vicinity in the sky. We do this by examining their \textit{Kepler} target pixel files (TPFs). TPFs show a `postage stamp' image of pixels around \textit{Kepler} targets, with one image at every cadence ($\Delta T \simeq 30$min.) and each pixel spanning 4 arcseconds across the sky. These files are available to download from the Mikulski Archive for Space Telescopes (MAST)\footnote{\href{http://archive.stsci.edu/kepler/}{http://archive.stsci.edu/kepler/}}, which we retrieve and access using the {\fontfamily{qcr}\selectfont{lightkurve}}\footnote{\href{https://lightkurve.keplerscience.org/}{https://lightkurve.keplerscience.org/}} Python package \citep{lightkurve}. 

A pixel image from the TPF provides a `flux map' (Figure \ref{Maps_Known}, right), which describes light intensities in regions surrounding a \textit{Kepler} target for a given observation timestamp. Since the full set of pixel images from the TPF provides a light curve for each individual pixel, we create power spectra for each pixel. First, we prepare the light curves by passing them through a high-pass boxcar filter of 20-day length and perform 3$\sigma$ clipping to remove outliers before taking its Fourier transform, similar to the approach by \citet{Colman_2017}. This effectively produces a `power spectrum map' (Figure \ref{Maps_Known}, left) to complement the flux map. To further aid the identification of the polluting star, we overlay a sky map on the flux map, where we plot and compare the celestial coordinates of the main \textit{Kepler} target with that of other nearby known stars from various catalogs.

Using the flux and power spectrum maps, we can identify the pixels that have high incident fluxes and show oscillations, which we then supplement with spatial information from the star map. Such an approach allows us to easily find potential candidates for the polluting star. Because the orientation of the telescope during the mission changes by $90^\circ$ every quarter, we examine each quarter of pixel data separately. In the following, we describe how we search and identify the polluting giants from different survey catalogs, in decreasing order of priority.
\subsection{Identification of polluting giants}
\subsubsection{\textit{Kepler} targets}\label{Kepler_target_id}

\begin{figure*}
\centering
	\includegraphics[width=\linewidth]{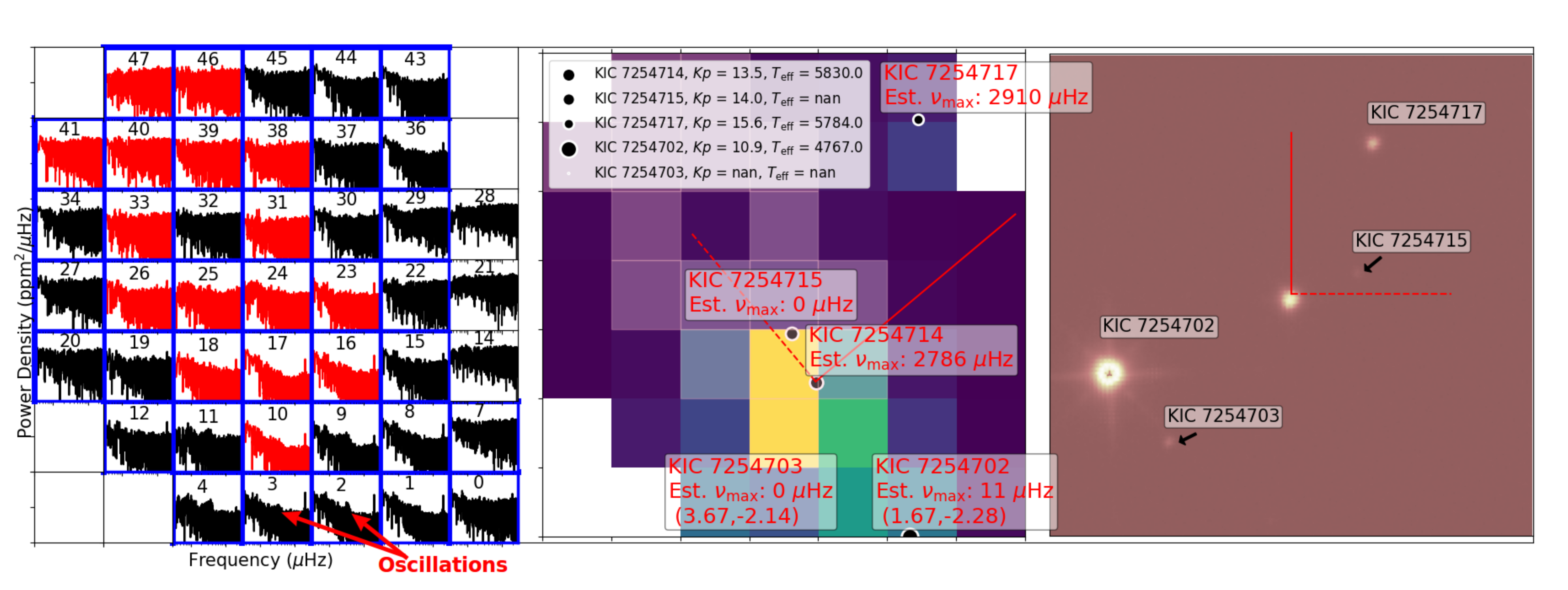}
	\caption{Target Pixel File for Quarter 4 of the blended target KIC 7254714, with its power spectrum map (left) and flux/star maps(center). Additionally, a 45 x 45 arcsecond \textit{Kepler} field image cutout from the UKIRT WFCAM survey centered on the blended target is displayed on the right. To indicate scale, compass arms are drawn on both the star map and the field image and both are 15 arcseconds across. The solid line points north, while the dashed line points east. $\nu_{\mathrm{max,sc}} = 0$ is assigned to stars without $T_{\mathrm{eff}}$ or $R$ values. The polluting star is identified as KIC 7254702, which is a very bright target such that its oscillation signal (as indicated in pixels 2 and 3 in the power spectrum map) can be detected across many pixels. The use of the UKIRT image allows us to visually identify bright polluting giants beyond our search range for non-\textit{Kepler} targets (2 pixels beyond the TPF).}

	\label{Maps_UKIRT}
\end{figure*}

We first attempt to identify the polluting giants with targets in the KSPC. In the power spectrum map of each suspected blended target, we initially identify which pixels contain the polluting red giant oscillation signal and determine if these pixels correspond to high flux values in the flux map. This allows us to approximately locate the position of the polluting star in the sky with respect to the blended target.

Next, we plot all other \textit{Kepler} targets that can be seen within the TPF on the sky map. In other words, we find all other targets with celestial coordinates within an area of sky centered around the blended target, with the area boundary determined by the dimensions of the TPF pixel image. Because light from a star can fall onto several different pixels due to the point-spread function of the \textit{Kepler} telescope \citep{Bryson_2010}, in some scenarios the polluting star may be located outside the TPF bounds. Thus, we extend our search boundary to also include stars that are 5 pixels (20 arcseconds) beyond each boundary of the TPF. 

For every star we locate in the vicinity of a blended target, we calculate $\nu_{\mathrm{max,sc}}$, which is an estimate the star's frequency at maximum power using the following scaling relation \citep{Brown_1991,Kjeldsen_1995}:

\begin{equation}
\frac{\nu_{\mathrm{max,sc}}}{\nu_{\mathrm{max\odot}}} \simeq \bigg(\frac{M}{M_{\odot}}\bigg)\bigg(\frac{R}{R_\odot}\bigg)^{-2}\bigg( \frac{T_{\mathrm{eff}}}{T_{\mathrm{eff\odot}}} \bigg)^{-1/2},
\label{scaling_relation}
\end{equation}

\noindent where $M$, $R$, and $T_{\mathrm{eff}}$ are the stellar mass, radius, and effective temperature, respectively, with $\nu_{\mathrm{max\odot}} = 3050$ $\mu$Hz and $T_{\mathrm{eff\odot}} = 5772$K. For our estimates, we adopt $T_{\mathrm{eff}}$ and $R$ values from B18. For all our estimates, we use a constant value of $M=1.2M_{\odot}$, which is approximately the median stellar mass of \textit{Kepler} red giants and is a good estimate to within a factor of two for the vast majority of the giants (see \citealt{Yu_2018}, their Figure 5a). While $\nu_{\mathrm{max,sc}}$ is only a rough estimate of $\nu_{\mathrm{max}}$, it will suffice for the purpose of selecting polluting candidates. We select the polluting star by identifying a neighbouring star that satisfies the following requirements:

\begin{enumerate} \label{enumer}
 \item The potential candidate is located at a position on the sky map that coincides with the location of the polluting star as inferred from the flux and power spectrum maps. \label{Point_One}
 \item $\nu_{\mathrm{max,sc}}$ of the potential candidate (Equation \ref{scaling_relation}) matches the $\nu_{\mathrm{max}}$ as measured by the deep learning regressor within 2$\sigma$. This can also be confirmed by visually identifying $\nu_{\mathrm{max}}$ from the power spectrum maps. An example of this selection process is shown in Figure \ref{Maps_Known}.\label{Point_Two}
\end{enumerate}

\subsubsection{\textit{Kepler} Input Catalog (KIC) stars} \label{KIC_id}

\begin{figure}
	\centering
	\includegraphics[width=\linewidth]{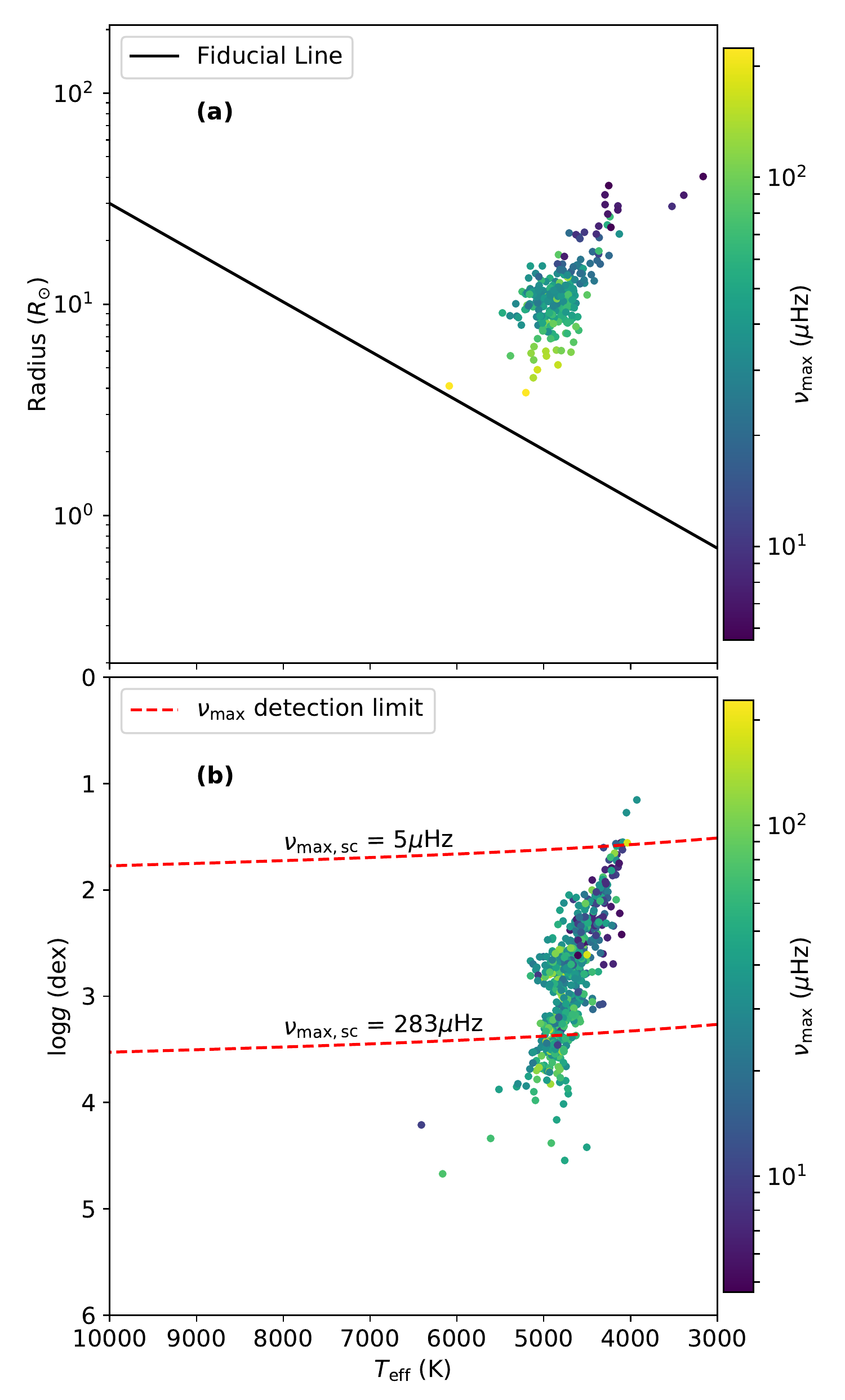}
	\caption{(a) 293 polluting giants that are known \textit{Kepler} targets, with $R$ and $T_\mathrm{eff}$ values adopted from the B18 catalog. The drawn fiducial line is the same as in Figure \ref{pred_gaia_rad}. (b) 562 polluting giants that are non-\textit{Kepler} targets in the KIC, with $\log g$ and $T_\mathrm{eff}$ values adopted directly from the KIC. 25 stars do not have $\log g$ values from the KIC and hence are not plotted. The dashed red lines are lines of constant $\nu_{\mathrm{max,sc}}$ and is a rough guide to indicate the detection limit of our method.}
	\label{BlendVerification}
\end{figure}

If there are no \textit{Kepler} targets that are suitable candidates for the polluting star, we next search for candidates within the KIC. Such stars typically have stellar parameters in the KIC but do not have light curves as they were not targeted by \textit{Kepler}. We perform a similar analysis as described in Section \ref{Kepler_target_id}, except that here we restrict our search boundary to 2 pixels (8 arcseconds) beyond each TPF boundary. This is because searching for a polluting candidate star beyond the TPF becomes increasingly difficult with larger search boundaries due to the high density of stars now being included from the KIC.

This time, we adopt $R$ and $T_\mathrm{eff}$ values directly from the KIC. Because not all entries in the KIC have these values measured, three scenarios are possible:
\begin{enumerate}
\item The polluting candidate has both $T_\mathrm{eff}$ and $R$ listed, allowing a determination of  $\nu_{\mathrm{max,sc}}$ using Equation \ref{scaling_relation}, hence allowing us to match $\nu_{\mathrm{max,sc}}$ with $\nu_{\mathrm{max}}$ values from the deep learning regressor and $\nu_{\mathrm{max}}$ observed from the power spectrum maps.
\item The polluting candidate only has $T_\mathrm{eff}$ measured. We identify stars with $3000$K $\leq T_\mathrm{eff}\leq6000$K as more likely candidates because they have the expected $T_\mathrm{eff}$ for a red giant. The availability of a $T_\mathrm{eff}$ value also indicates the presence of multi-band photometry for this star, hence it is typically not too faint.
\item The polluting candidate has neither $T_\mathrm{eff}$ nor $R$ listed. In this scenario, we are only able to rely on the spatial requirement \ref{Point_One} in Section \ref{enumer}.
\end{enumerate}

For uncommon scenarios where identifying the polluting star is difficult, we also compare the flux, power spectra, and sky maps from each TPF with higher resolution images of the same area of sky using \textit{Kepler} field image cutouts from the UKIRT WFCAM (UK Infrared Telescope Wide Field Camera) survey \citep{Lawrence_2007}. An example of this is shown in Figure \ref{Maps_UKIRT}.

\subsubsection{Gaia targets within the \textit{Kepler} field}
For polluting giants that are not identified as \textit{Kepler} targets or as entries in the KIC, we next identify them using data from the Gaia mission \citep{Gaia_mission}, where we search across 10.3 million Gaia targets from the Gaia DR2 catalog \citep{Gaia_DR2} that lie within $8^{\circ}$ from the center of the \textit{Kepler} field. We use a similar approach to that in Section \ref{KIC_id}, except that we adopt $T_\mathrm{eff}$ and $R$ values from the Gaia DR2 catalog.
\begin{table} 
			\centering
			\caption{A list of blended targets, each assigned with the identity of its polluting star. Flag `0' indicates that the polluting star is a \textit{Kepler} target, flag `1' indicates that the polluting star is a non-\textit{Kepler} target in the KIC, flag `2' indicates that the polluting star is only identified in the Gaia DR2 catalog (for which the 19-digit Gaia source ID is listed instead), while flag `3' indicates an unidentified polluting star. The full version of this table is available in a machine-readable format in the online journal, with a portion shown here for guidance regarding its form and content.}
			\label{Blend_Table}
			\begin{threeparttable}
				\begin{tabular}{|c|c|c|}
					\hline
					Blended Target ID& Polluting Star ID & Flag\\
					\hline
					7102071&7102068&0\\
					7207089&7124606&1\\
                    7211526&7211529&1\\
                    7222444&7222454\tnote{*}&1\\
                    7797928&-&3\\
                    7811846&7811847&0\\
                    7812628&7812622&1\\
                    7881258&7881261&0\\
					7984047&-&3\\
					8016196&2105691730722492544\tnote{*}&2\\
					...&...&...\\
					\hline
					
				\end{tabular}
                \begin{tablenotes}
                \item[*] No $T_{\mathrm{eff}}$ or $R$ available for this star
				\end{tablenotes}
			\end{threeparttable}
		\end{table}
		
\begin{figure*}
\centering
	\includegraphics[width=0.9\linewidth]{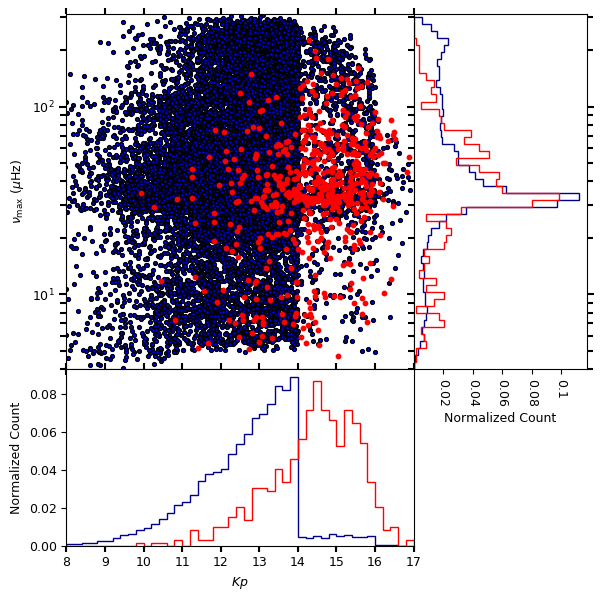}
	\caption{Comparison of $\nu_{\mathrm{max}}$ vs. $Kp$ for 21,005 giants above the fiducial line in Figure \ref{pred_gaia_rad}a (blue) and for the 587 polluting non-\textit{Kepler} targets in the KIC (red). The distributions of the red and blue dots are shown as histograms in $Kp$ and $\nu_{\mathrm{max}}$.}
	\label{Above_Below_Histograms}
\end{figure*}

\subsection{Analysis of polluting giants}\label{det_below_fiducial}
From the 909 dwarf/subgiant stars with giant oscillations, we identify that 293 have polluting giants that are \textit{Kepler} targets (hence with previously known light curves on MAST). Hence, the remaining 616 are newly identified oscillating red giants, with 587 of them found to be non-\textit{Kepler} targets in the KIC, while 14 are only identified as Gaia targets, and the remaining 15 are unidentified. For brevity, we refer to the  sample of the 616 newly identified oscillating giants as the `new giants'.

First, we verify that all our identified polluting giants are indeed red giants. In Figure \ref{BlendVerification}a, we plot the stellar parameters of the 293 polluting giants that are known \textit{Kepler} targets. As expected, they all appear above the fiducial line and hence are red giants. In Figure \ref{BlendVerification}b, we do the same for the 587 polluting non-\textit{Kepler} targets that are in the KIC, except that we plot the $\log g$ and $T_{\mathrm{eff}}$ values from the KIC. We find that most have $5\mu$Hz $\apprle \nu_{\mathrm{max,sc}} \apprle 283\mu$Hz as expected for giants detectable by our classifier . However, there is a significant number of stars with $\nu_{\mathrm{max,sc}} > 283\mu$Hz. This can be explained by the fact that $\log g$ values from the KIC are relatively uncertain and typically overestimated by about 0.5 dex \citep{KIC}. Alternatively, the detected oscillations from these stars may be aliased counterparts of super-Nyquist modes as discussed in Section \ref{estimating_surface_gravity}.

To investigate the properties of the new giants, we compare the $Kp$ and $\nu_{\mathrm{max}}$ distributions of the 587 polluting non-\textit{Kepler} targets that are in the KIC with the positive detections above the fiducial line. The result of this is shown in Figure \ref{Above_Below_Histograms}. For $\nu_{\mathrm{max}}\apprge 100\mu$Hz, the number of new red giants (red) decreases more rapidly with $\nu_{\mathrm{max}}$ compared to the detections above the fiducial line (blue), which is expected because each photometric aperture that measures the flux from these new red giants is not centered around them. As a result, these new giants are measured `indirectly', causing them to have lower signal-to-noise levels as compared to \textit{Kepler} targets with apertures centered around them. As expected, compared to \textit{Kepler} targets, these new serendipitous giants are generally given less confident probabilities by the classifier as shown in Appendix \ref{Blends_Probability}.

As a whole, the $Kp$ distribution of these new giants differ significantly from the distribution for targeted giants, which shows a cut-off at $Kp\simeq 14$ from their selection function \citep{Batalha_2010,Sharma_2016}. Specifically, it can be seen that a significant fraction of the new giants are fainter than the majority of the targeted giants. They therefore potentially represent a population of distant stars similar to the giants residing in the halo of the Galaxy as identified by \citet{Mathur_2016}. We find that our sample of new red giants is widely distributed spatially within the \textit{Kepler} field (see Appendix \ref{Blends_Coordinates}) and has no overlap with the sample by \citet{Mathur_2016}, hence it will prove to be valuable for galactic archaeology.

\begin{figure}
	\centering
	\includegraphics[width=\linewidth]{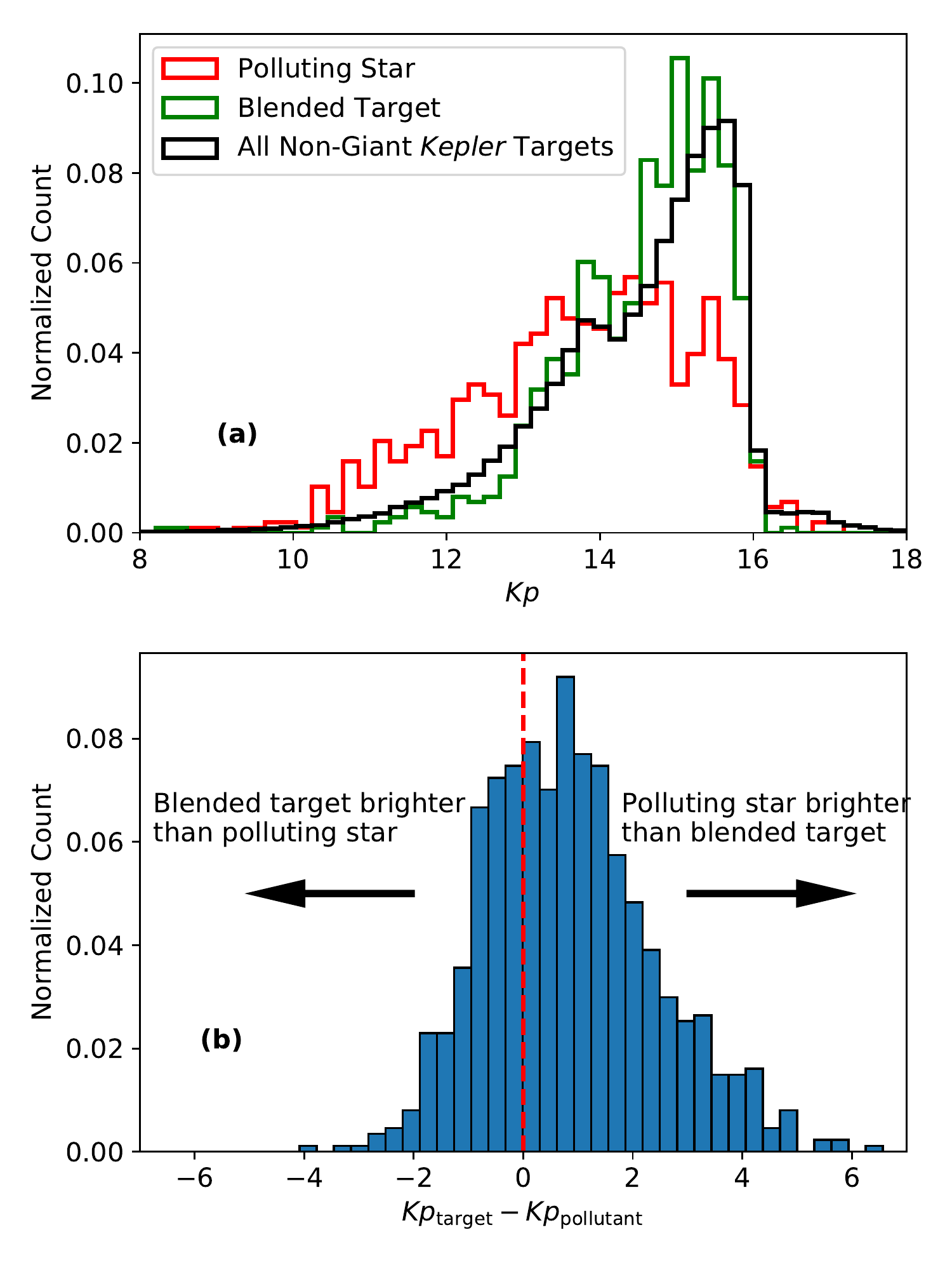}
	\caption{(a) Comparison of \textit{Kp} distributions between blended targets (green), which are the stars below the fiducial line in Figure \ref{pred_gaia_rad}a, and their polluting giants (red), which are red giants neighbouring the blended targets that cause red giant solar-like oscillations to be seen in the power spectrum of the blended targets. The $Kp$ distribution for non-giant \textit{Kepler} targets (black) is also shown for comparison. (b) $Kp$ differences between a blended target ($Kp_{\mathrm{target}}$) and its identified polluting star ($Kp_{\mathrm{pollutant}}$). The dashed red line indicates the zero point, where both stars' magnitudes are equal.}
	\label{Polluted_Pollutant_Hist}
\end{figure}

Besides polluting giants with entries in the KIC, our new giants also include 14 that are only identified as Gaia targets. In most of these scenarios, the flux and power spectrum maps indicate that the polluting star should be located within an adjacent pixel to the blended target, where we indeed find a nearby Gaia target within such an area. However, all Gaia targets that we identify have no measured $T_{\mathrm{eff}}$ from the Gaia DR2 catalog, such that we can only determine polluting star candidates from their location on the star map relative to the blended target. This requirement alone is not strong evidence towards the identity of the actual polluting star. Hence, it is possible that the polluting giants for these targets may instead be from unresolved binary members or chance alignments from other background stars.

We are unable to identify the polluting giants for 15 dwarfs/subgiants that show giant oscillations. Eleven of these are marginal detections where most have been observed for only a few quarters. We infer that the remaining 4 may have unresolved polluting giants, because their TPFs indicate polluting giants that are close to the blended target, but no such star can be identified from either the KIC or the Gaia DR2. In Table \ref{Blend_Table}, we tabulate the 909 blended targets along with their corresponding identified polluting star. The $\nu_{\mathrm{max}}$ for each polluting star can be found by matching the blended target ID in Table \ref{Blend_Table} with the KIC ID in Table \ref{Detection_Table}.

Finally, we investigate the difference in $Kp$ distributions between all polluting giants with entries in the KIC (red) and the \textit{Kepler} targets that they pollute (green) as shown in Figure \ref{Polluted_Pollutant_Hist}a. We see that these two distributions are significantly different from one another, which we can better quantify by examining the distribution of their $Kp$ differences as shown in Figure \ref{Polluted_Pollutant_Hist}b. On average, the polluting star is of comparable or brighter magnitude compared to the blended target, which is an expected result. Interestingly, there are also polluting giants that are 2$\sim$4 magnitudes fainter than the blended target that are still detected. This can be informative about the $Kp$ detection limit for neighbouring stars using our custom aperture. While further investigations into these distributions are beyond the scope of this paper, the sample of blended targets and their polluting giants will nonetheless be of significant interest towards modeling populations of giants in the \textit{Kepler} field.

\section{Discussion and Conclusion}\label{Discussion}
From all $\sim$197,000 \textit{Kepler} stars observed in long-cadence, we detected a total of 21,914 stars showing solar-like oscillations, yielding the current largest list of stars showing solar-like oscillations that will provide many more targets of interest for various asteroseismic and Galactic archaeology analyses. We also predict $\nu_{\mathrm{max}}$ for each detection, which will provide useful prior values for more precise measurements using asteroseismic model-fitting pipelines. In addition, our $\nu_{\mathrm{max}}$ values provide $\log(g)$ estimates, for which $\sim$88\% are good to within 0.05 dex, which is still superior to typical spectroscopic $\log(g)$ determinations. As a follow up, it will be useful to compare our $\nu_{\mathrm{max}}$ values with measurements from even more classical seismic pipelines other than the one used here (A2Z), as well as with the FliPer metric \citep{Bugnet_2018}, or statistical tests on the power spectrum \citep{Bell_2018}.

From our list of 21,914 detections, we identified 21,005 \textit{Kepler} targets with Gaia-derived radii representative of red giants. Even though we do not account for giants with $R\apprge40R_{\odot}$, our number of detections is within estimates of the total number of red giants ($\sim$21,000) as reported by B18 because we also detect giant oscillations in stars with $R\apprle10R_{\odot}$ that they classified as subgiants (see Figure \ref{pred_gaia_rad}). We also found 1,671 stars with $R\leq40R_\odot$ that are predicted as non-detections but have been classified as red giants by B18. We infer that these are potentially stars experiencing suppression of their oscillations due to strong dynamical interactions such as rapid rotation in close binary systems \citep{Gaulme_2014,Tayar_2015}.

By observing the $\nu_{\mathrm{max}}$-$Kp$ distribution of the 21,005 detected giants, we obtained an empirical estimate for the detection limit for solar-like oscillations. Because our detection of oscillations are based on qualitative `visual' methods, our empirical detection limit should provide a reasonable estimate to the detectability of oscillations that can be derived using other methods such as classical seismic pipelines.

We also compare our detections with predictions from a synthetic model of the Milky Way by Galaxia, where we find a good agreement of the distributions of the giants in $(Kp,\log g_{\rm kic})$ and $(\nu_{\rm max},Kp)$ space. Furthermore, a comparison of the number of observed and simulated oscillating giants indicate that our detections are $\sim$99\% complete for stars with $10\mu$Hz $ < \nu_{\mathrm{max}} < 270\mu$Hz. This level of completeness decreases to $\sim$97\% for stars with $3\mu$Hz $ < \nu_{\mathrm{max}} < 270\mu$Hz as a result of the limitation of our deep learning classifier at low $\nu_{\mathrm{max}}$.

Our study has used custom photometric apertures that are optimized for light curve stability on longer time scales, hence they are generally larger than the apertures by the \textit{Kepler} Science Processing Pipeline by a factor of three (cf. Figures \ref{Maps_Known} and \ref{Maps_UKIRT}). The resulting extra white noise in the light curves is, however, small enough such that we expect only the detection of super-Nyquist stars with very low SNR to be affected. This is because larger aperture sizes are required for fainter stars. Thus, our use of custom apertures has not significantly affected most of our detections, and has been beneficial by enabling us to identify as many as 909 serendipitous detections from blended \textit{Kepler} targets. These 909 serendipitous detections show oscillations indicative of giants but have Gaia-derived radii representative of dwarf stars. Using the Target Pixel Files for such stars, we determined that most were blended targets as a result of using a larger photometric aperture. We identified the polluting giants for most of the 909 blended targets, where 293 polluters are \textit{Kepler} targets, while 587 are non-\textit{Kepler} targets that are, however, still listed in the KIC. For the non-\textit{Kepler} polluting giants that are identified in the KIC, we found that their $Kp$ distribution differs significantly from the distribution of \textit{Kepler} red giant targets, with a large fraction of polluting giants having $Kp>14$. We also identified potential polluting giants found only in the Gaia DR2 catalog for 14 blended targets, while the identification for the 15 remaining blended targets remains uncertain. We suggest that stars that show clear detections among these 15 stars may have unresolved nearby polluting giants. 

Altogether, this forms a total of 616 out of 909 blended targets that were not targeted by \textit{Kepler}. This opens up opportunities for performing asteroseismology on oscillating red giants without previous light curves of their own, which will be valuable additions to both stellar population studies and Galactic archaeology. The addition of over 600 new giants, from which many have been observed for $\sim4$ years is also significant for asteroseismology, since \textit{Kepler} giants are likely to have the longest time series for giants in the foreseeable future.

Currently, our method is limited to stars with $\nu_{\mathrm{max}} \apprge 4\mu$Hz, such that we can only automatically identify oscillating red giants with $R\apprle40R_{\odot}$. With the greater availability of highly luminous giants that are analysed seismically (Yu et al. in preparation), we will explore extending our method to lower frequencies. In Section \ref{classifier_limitation}, we addressed additional classifier limitations that arose as a result of generalizing our K2-optimized classifier towards the full \textit{Kepler} long cadence dataset. Despite these limitations, our current classifier significantly narrowed the search space for detecting oscillating giants, resulting in only a small subset of targets that required follow-up visual inspection. This is evidenced by a total of 21,721 out of the final 21,914 oscillating giants that have been detected by the deep learning classifier, with the remaining giants manually included after comparing with classical seismic pipeline results and visual inspection.

The detected oscillating giants from this study will be part of a new training set based on the full \textit{Kepler} long cadence data that will be much larger and more robust than our current training set. Our next implementation of the deep learning classifier that trains with this new set will hence benefit from a much stronger capability to distinguish solar-like oscillations across a wider range of stellar variability compared to the classifier in this study. This will be very useful for analyzing the upcoming data from the Transiting Exoplanet Survey Satellite (TESS) \citep{TESS}, where we intend to run our deep learning classifier in a fully-automated fashion to classify millions of stars from the mission.

	\section*{Acknowledgements}
Funding for this Discovery mission is provided by NASA's Science mission Directorate. We thank the entire \textit{Kepler} team without whom this investigation would not be possible. D.S. is the recipient of an Australian Research Council Future Fellowship (project number FT1400147). R.A.G. acknowledges the support from CNES. S.M acknowledges support from NASA grant NNX15AF13G, NSF grant AST-1411685, and the Ramon y Cajal fellowship number RYC-2015-17697. I.L.C. acknowledges scholarship support from the University of Sydney. We would like to thank Nicholas Barbara and Timothy Bedding for providing us with a list of variable stars that helped to validate a number of detections in this study. We also thank the group at the University of Sydney for fruitful discussions. Finally, we gratefully acknowledge the support of NVIDIA Corporation with the donation of the Titan Xp GPU used for this research.




\bibliographystyle{mnras}
\bibliography{bibi4} 




\appendix

\section{Non-Detections with Gaia-Derived Red Giant Radii}\label{RG-NonDet}
In Table \ref{RG_Non-Det_Table}, we present a list of 1,671 \textit{Kepler} targets (available online) with $R\leq40R_\odot$ that were classified as red giants by \citet{Berger_2018} but are not predicted to show red giant solar-like oscillations in this study. As discussed in Section \ref{Discussion}, oscillation modes in these stars are potentially suppressed, which can be caused by strong dynamical interactions such as rapid rotation in close binary systems \citep{Gaulme_2014,Tayar_2015}.
\begin{table} 
			\centering
			\caption{A list of \textit{Kepler} targets with radii, $R$, smaller than $40R_\odot$ that are predicted to not show giant oscillations but are classified as red giants by \citet{Berger_2018}. The stars are identified by the \textit{Kepler} Input Catalog (KIC) ID, with radii adopted from the catalog from \citet{Berger_2018}. The full version of this table is available in a machine-readable format in the online journal, with a portion shown here for guidance regarding its form and content.}
			\label{RG_Non-Det_Table}
			\begin{threeparttable}
				\begin{tabular}{|c|c|}
					\hline
					KIC ID & Radius ($R_\odot$)\\
					\hline
					1995430& 5.465\\
                    1995908& 3.505\\
                    2010161& 4.592\\
                    2011145& 35.833\\
                    2011428& 10.706\\
                    2011545& 4.451\\
                    2011870& 33.207\\
                    2011908& 3.642\\
                    2011945& 5.451\\
					...&...\\
					\hline
					
				\end{tabular}

			\end{threeparttable}
		\end{table}

\section{Double blended target KIC 5824221}\label{Appendix_Double_Blend}
\begin{figure*}
\centering
	\includegraphics[width=\linewidth]{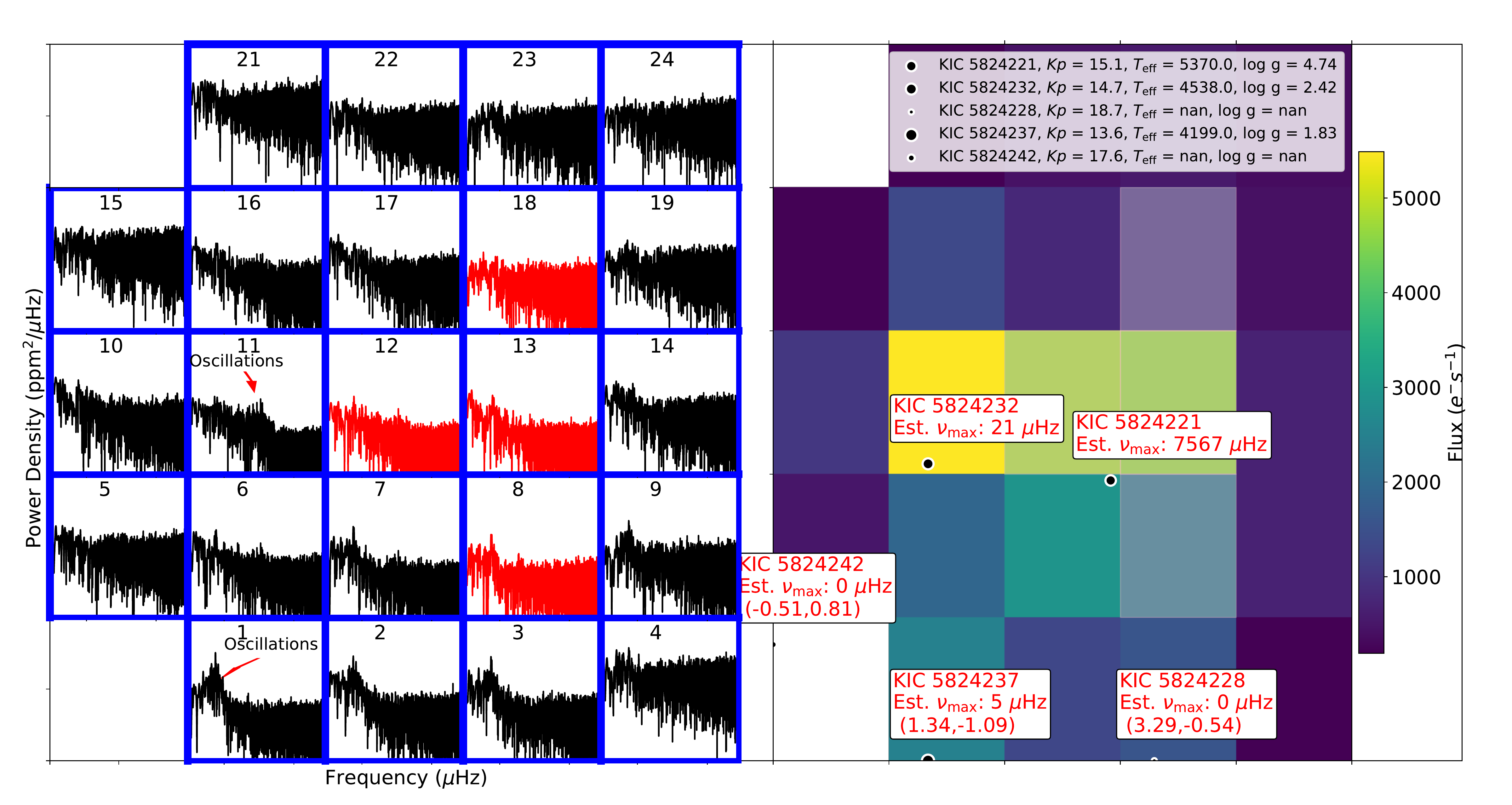}
	\caption{Target Pixel File for Quarter 3 of the double blended target KIC 5824221 with its power spectrum map (left) and its flux/star map (right) . $\nu_{\mathrm{max,sc}} = 0$ is assigned to stars without $T_{\mathrm{eff}}$ or $R$ values from the B18 catalog. The source of the polluting signals can be seen mainly in pixels 1 and 11, which agrees with the position and $\nu_{\mathrm{max}}$ for the stars KIC 5824237 and KIC 5824232, respectively. The format for the Figure is identical to that of Figure \ref{Maps_Known}.}

	\label{Double_Blend_Maps}
\end{figure*}

In Figure \ref{Double_Blend_Maps}, we demonstrate an example of the identification of the polluting star for the blended target KIC 5824221. The power spectrum for this blended target is shown in Figure \ref{ExampleBlend}.
The two power excesses arises as a result of pollution from two nearby red giants, which we identify to be KIC 5824237 and KIC 5824228.

\section{Detection Probability of Blends from Classifier} \label{Blends_Probability}
\begin{figure}
\centering
	\includegraphics[width=1.\linewidth]{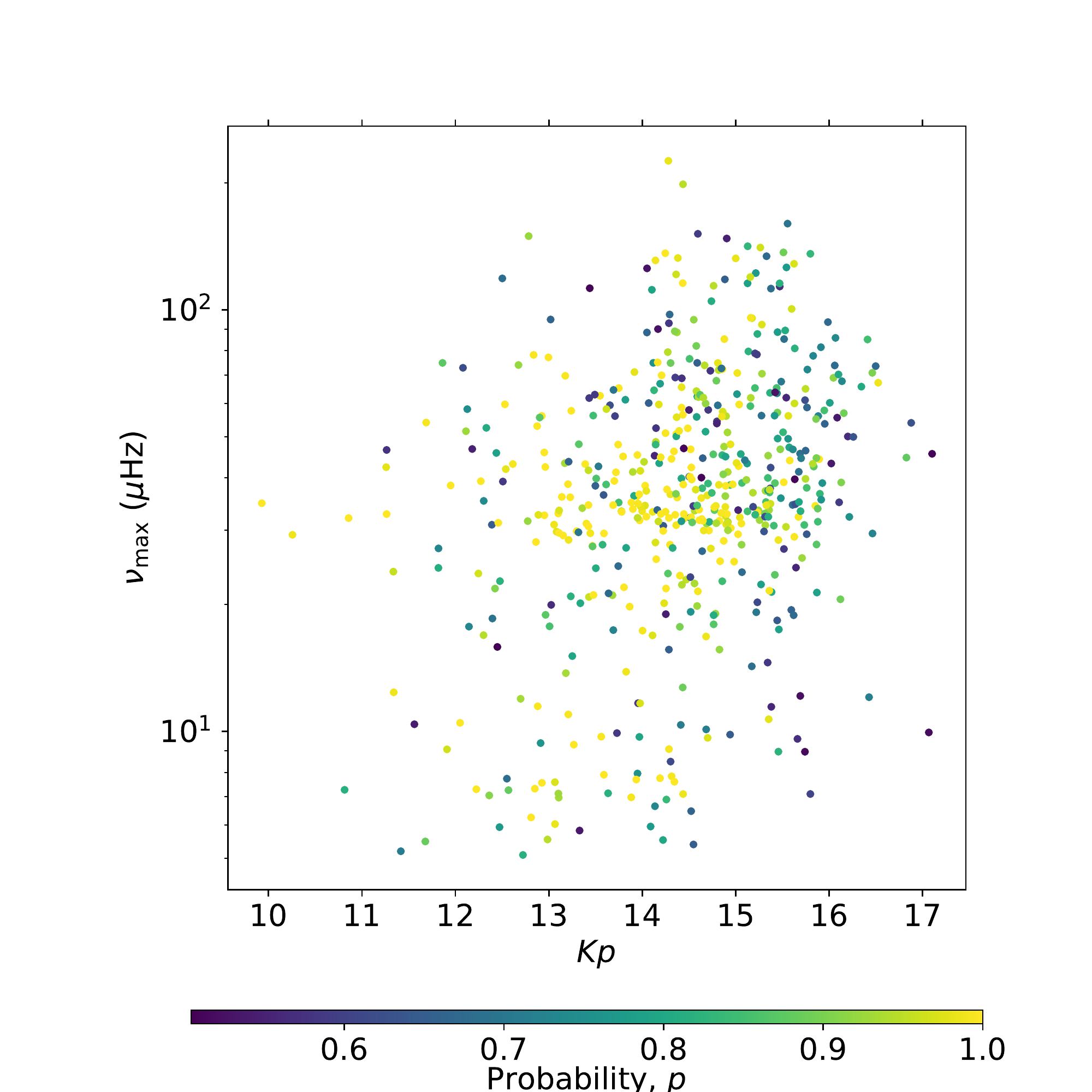}
	\caption{$Kp-\nu_{\mathrm{max}}$ distribution of the 587 polluting non-\textit{Kepler} targets that are in the KIC, colour-coded by classifier probability.}

	\label{Blends_Probability_Image}
\end{figure}

In Figure \ref{Blends_Probability_Image}, we show the classifier probabilities of the 587 polluting non-\textit{Kepler} targets that are in the KIC. Compared to the distribution of probabilities shown in Figure \ref{posdet_nonblend_hist}, the probabilities shown here are generally not as confident, shown by a larger occurrence of dark-coloured points. The reason for this is because they have been detected `indirectly' (as discussed in Section \ref{det_below_fiducial}) and hence their power spectra contain higher noise levels than that of main \textit{Kepler} targets.

\section{Distribution on Identified Polluting Giants in \textit{Kepler} Field of View (FOV)}\label{Blends_Coordinates}
\begin{figure}
\centering
	\includegraphics[width=1.\linewidth]{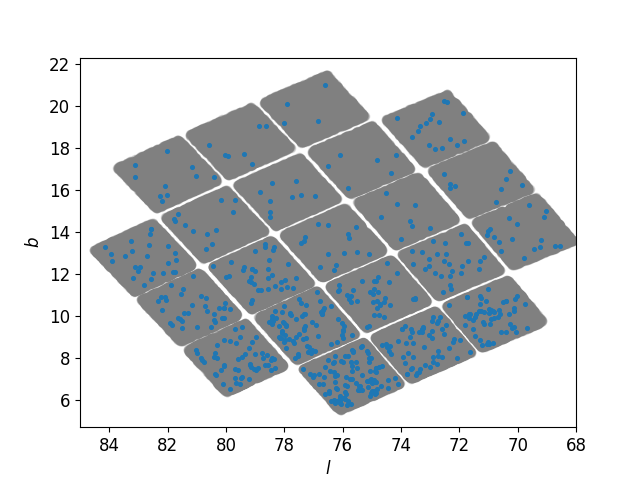}
	\caption{The 587 polluting non-\textit{Kepler} targets that are in the KIC, plotted over the \textit{Kepler} FOV in Galactic coordinates. As expected, more pollution occurs near the Galactic plane (low $b$).}

	\label{Blend_Coords}
\end{figure}

In Figure \ref{Blend_Coords}, we plot the Galactic coordinates of the 587 polluting non-\textit{Kepler} targets in the KIC over the \textit{Kepler} FOV, where it can be seen that they are widely distributed, with a tendency of more stars closer to the Galactic plane (lower $b$) as expected.

\bsp	
\label{lastpage}
\end{document}